\documentclass[12pt]{article}
\usepackage{amsmath, amssymb}
\usepackage{epsfig}
\newcommand{\postscript}[2] {\setlength{\epsfxsize}{#2\hsize} \centerline{\epsfbox{#1}}}
\begin{document}
% Journal identifier can be put here if required, e.g.

\title{A new topology on the space of Lorentzian metrics on a fixed manifold}
\maketitle
\author{Johan Noldus, Faculty of Applied Sciences, department of mathematical analysis,  University of Gent,  Galglaan 2, 9000~Gent, Belgium}

\begin{abstract}
We give a covariant definition of closeness between (time
oriented) Lorentzian metrics on a manifold M, using a family of
functions which measure the difference in volume form on one hand
and the difference in causal structure relative to a volume scale
on the other hand.  These functions will distinguish two geometric
properties of the Alexandrov sets $ A(p,q), \tilde{A} (p,q) $
relative to two space time points $q$ and $p$ and metrics $g$ and
$ \tilde{g} $.  It will be shown that this family generates
uniformities and consequently a topology on the space of
Lorentzian metrics which is Hausdorff when restricted to strongly
causal metrics. This family of functions will depend on parameters
for a volume scale, a length scale (relative to the volume scale)
and an index which labels a submanifold with compact closure of
the given manifold M.
\end{abstract}


%\maketitle

%%%%%%%%%%%%%%%%%%%%%%%%%%%%%%%%%%%%%%%%%%%%%%%%%%%%%%%%%%%%%%%%%%%%%%%%%%%%%%%%%%%%%%%%%%%%%%%%
\section{Introduction}
In this paper we follow the convention that M is a Hausdorff,
paracompact, oriented $ C^{ \infty } \quad d+1$ - dimensional
manifold. When $M$ is compact , we shall assume it has a boundary
otherwise the chronology violating set wouldn't be empty, ie.
there exist timelike closed curves ~\cite{Hawking1}. In the other
case $M$ is assumed to be a manifold without boundary. We will
first consider compact space-times for technical reasons, later on
we will propose a generalization to the non compact case.
\\ There exist already a great deal of topologies, most of
them however do not use the specific properties which emerge from
the symmetry and signature of a Lorentzian metric.  An example is
the $ W^{k} $ compact - open topology which is defined by the open
sets : $$ B_{ X \epsilon }(g) = \left\{ \tilde{g} \mid {\parallel
g - \tilde{g}, X \parallel}_{ W^{k}} < \epsilon \right\} $$ and
$$ {\parallel g - \tilde{g},X \parallel}_{ W^{k}} = \sqrt{ \sum_{ p = 0}^{k} { \int_{X}{
{ \parallel D^{p} ( g -\tilde{g} ) \parallel }^{2} d \sigma }}}
$$ $d \sigma $ is the volume form induced by a reference
Riemannian metric $h$, $ X$ is an open submanifold with compact
closure in M, $D$ is a reference derivative operator and $$
\parallel D^{p} ( g -\tilde{g} )\parallel = \sqrt{ (g -\tilde{g}
)_{ab ; c_{1}...c_{p}} h^{ad} h^{be}
h^{c_{1}f_{1}}...h^{c_{p}f_{p}}(g -\tilde{g} )_{de ;f_{1}...f_{p}}
} $$ It has been proven that this topology is independent of the
chosen Riemannian metric and derivative operator. This topology is
more local than the $ C^{k} $ open topology which demands that the
difference of the metrics and their derivatives becomes uniformly
small.   The $ W^{k} $ topology is therefore coarser than the
uniform convergence topology. \\ A more "physical" approach was
taken by L. Bombelli and R. D. Sorkin, using the fact that causal
structure and conformal structure are the same when the Lorentzian
metrics are future and past distinguishing.  In that filosophy
they defined a set of functions which compare the volume elements
and causal structures of two metrics $g$, $ \tilde{g} $
separately. They prove that this topology is Hausdorff when
restricted to $ C^{2}$ future and past distinguishing metrics. By
definition this means that for every point $p$ there exists a
unique maximal geodesic with starting point $p$ and initial
direction $ X_{p} \in T_{p}M $. Moreover the geodesic depends
continuously on $p$ and $X_{p}$ in the sense that $ \forall
(p,X_{p}) \in TM \quad \forall V \subset M $, V an open
neighborhood of $\exp(p,X_{p}) \quad \exists U$ open in $TM$ such
that $ \exp(q,X_{q}) \in V $ $\forall (q,X_{q}) \in U$ whenever $
\exp(q,X_{q}) $ is defined.
\\ Geometrically it seems reasonable that we do not interpret
perturbations of the form $ g + \delta g $, where $ \delta g$
results from an infinitesimal diffeomorphism, as genuine
perturbations. Therefore all "distance" functions should be fully
diffeomorphism invariant in the sense that $$ d(g, \tilde{g}) =
d(g, \psi^{*} \tilde{g}) \quad \forall \psi \in Diff(M) $$ In
~\cite{Bombelli,Bombelli2} one starts from functions which are
diagonally diffeomorphism invariant in the following sense:
$$  d(g, \tilde{g}) = d( \psi^{*} g, \psi^{*} \tilde{g})$$  $ \forall \psi \in Diff(M) $
as will be the case for my construction.  Later on it is argued in
~\cite{Bombelli,Bombelli2} that one can take the quotient $
\mathbb{L} (M)/Diff(M) $ in a well defined way, but one cannot
make any prediction anymore about the Hausdorff character of the
resulting topology.
\\
In this article we do not try to solve the problem of constructing
a fully diffeomorphism invariant metric topology.  We will try to
give a method which will solve the problem as good as possible
(within our knowledge) relying only upon contemporary functional
analytical methods.  There are two key concepts: on one hand there
is the choice of the topology of the diffeomorphism group (the
Schwartz topology) and on the other hand there is the concept of
amenability. A topological group $G$ is amenable if there exists a
positive translation invariant functional $A$ on the Banach space
of the bounded Borel measurable functions on $G$ such that
$A(1_{G} ) = 1$. The difficulty is captured in the concept of
measurability. The famous Banach Hausdorff Tarski paradox proves
the existence of a finitely additive measure on all subsets of $
\mathbb{R}^d$ when $ d \leq 2$ which is invariant under all
euclidian isometries.  It says moreover that such a measure does
not exist in higher dimensions.  Von Neumann remarked that this
curiosity is due to the structure of the isometry group and not to
the space it acts upon. He proved that $ SO(3) $ endowed with the
discrete topology contains a free group of two generators, and
therefore fails to be amenable, ie there does not exist a
translation invariant mean on the space of all bounded functions
of $ SO(3) $ ~\cite{Greenleaf,Pier}. However it is clear that $
SO(3) $ as a compact Lie group is amenable; the measurable sets
here are just the Borel sets.
\\
Suppose for the moment that the space-time manifold $M$ is compact
and define :
$$ d_{ \text{cau} }( g, \tilde{g}) = \int_{ M \times M} { \alpha_{g, \tilde{g}}(p,q) dV(p)
dV(q) }$$  where $$ \alpha_{g \tilde{g}}(p,q) =  \begin{cases}
\frac{ V( A(p,q) \bigtriangleup \tilde{A}(p,q)) }{ V( A(p,q) \cup \tilde{A}(p,q)) } & \text{if} \quad V( A(p,q) \cup \tilde{A}(p,q)) > 0  \\
 0 & \text{otherwise}
\end{cases} $$
$A(p,q) $ and $\tilde{A}(p,q)$ denote the Alexandrov sets with
respect to the metrics $ g$ and $ \tilde{g} $ respectively.  One
can define now a group action of $ Diff(M) $ on $ \mathbb{L}(M) $
by $ \psi \rightarrow \psi^{*} $ where $ \psi^{*} g $ is the pull
back of $ g $ with respect to $ \psi $. It is clear that $ d_{
\text{cau} }( g, \tilde{g} ) $ is diagonally diffeomorphism
invariant but in general
$$ d_{ \text{cau} }( g, \tilde{g} ) \neq d_{\text{cau}}( g, \psi^{*} \tilde{g})
$$  In the next chapter we will define functions $d_{ \text{vol}
}$ and $ d_{\text{geo}} $ which are also diagonally but not fully
diffeomophism invariant.  We will prove that these functions
generate a uniformity of countable basis; hence the corresponding
uniform topology is generated by a single pseudodistance $d$ which
is proven to be a distance on the space of the strongly causal
metrics.  This pseudodistance is evidently also diagonally
diffeomorphism invariant.
\\ Define now $ \forall g, \tilde{g} $ on $ Diff(M) $ the
following map:
$$ f_{g, \tilde{g}}^{ \text{cau}} : Diff(M) \rightarrow \mathbb{R} : \psi
\rightarrow d_{ \text{cau} }( g, \psi^{*} \tilde{g}) $$  The
logical question now is if for the Schwartz topology on  $ Diff
(M) $ and possible causal restrictions on $ g $ and $ \tilde{g} $
this map is Borel measurable.  Moreover one has to raise the
question if $Diff(M)$ is amenable and, if not,  whether one can
find a "large enough" subgroup which is?  Suppose for now that $ G
\subset Diff(M) $ is a maximal amenable subgroup and $A$ is a
invariant mean. Our strategy will be the following: first we prove
that $ d_{ \text{cau} }( g , \tilde{g} )$ is nonzero on the space
of \underline{strongly causal} $C^{2}$ metrics, then we prove that
when $g, \tilde{g} $ are globally hyperbolic the functions $
\alpha_{g \tilde{g}} $ and $ f_{g, \tilde{g}}^{ \text{cau}} $ are
continuous and hence measurable.  It will be obvious then that the
functions $ f_{g, \tilde{g} }^{\text{vol}} : \phi \rightarrow d_{
\text{vol}} ( g , \phi^{*} \tilde{g} ) $ and $ f_{g, \tilde{g}}^{
\text{ geo }} : \phi \rightarrow d_{ \text{geo} } ( g ,\phi^{*}
\tilde{g} ) $ are also continuous in the Schwartz topology.  This
implies that the function $f_{g , \tilde{g}} :  \psi \rightarrow
d(g , \psi^{*} \tilde{g} ) $ is continuous in the Schwartz
topology on $Diff(M)$. If we then define the function :
$$ \tilde{d} : ( g, \tilde{g} ) \rightarrow \begin{cases} A(f_{g , \tilde{g}} + f_{\tilde{g} ,g}) \quad \text{ when
} \quad \tilde{g} \neq \psi^{*} g
\quad \forall \psi \in Diff(M) \\
0 \quad \text{ otherwise } \end{cases}
$$ then $ \tilde{d} $ is a $G$ invariant pseudodistance
on the space of globally hyperbolic metrics.  $\tilde{d}$ is by
definition symmetric, so we have to prove the triangle inequality.
We have that $ \forall \phi , \psi \in G $: \begin{eqnarray*}  d(
g , \phi^{*} \tilde{g} ) & \leq & d( g , \psi^{*} \bar{g} ) + d(
\psi^{*} \bar{g} , \phi^{*} \tilde{g} ) \\ & \leq & d( g ,
\psi^{*} \bar{g} ) + d( \bar{g} , ( \phi \psi^{-1} )^{*} \tilde{g}
)
\end{eqnarray*} because $d$ is diagonally diffeomorphism
invariant.  Application of $A$ to this inequality (interpreted as
an inequality in functions of $ \phi $), using the right
invariance of $A$ and $ A(1) = 1 $, we get :
$$ A(f_{g, \tilde{g}}) \leq d( g ,
\psi^{*} \bar{g} ) + A (f_{ \bar{g} , \tilde{g} }) $$ Interpreting
this inequality as an inequality in functions of $ \psi $ we get
analogously :
$$ A(f_{g, \tilde{g} }) \leq A (f_{g, \bar{g}}) + A (f_{ \bar{g} , \tilde{g} }) $$
Symmetrization over $ g $ and $ \tilde{g} $ gives the desired
result.  We have made the assumption that $ g , \tilde{g} $ and $
\bar{g} $ were not diffeomorphism equivalent; the case where one
pair is equivalent is trivial.  The $G$ invariance of $\tilde{d}$
is a consequence of the left invariance of $A$. \\  We have no
argument to say that $\tilde{d}$ is a distance, but this issue is
discussed in greater detail later on. So when we restrict the
problem to globally hyperbolic metrics on a compact space - time,
the only open question which remains is the one concerning the
"size" of a maximal amenable subgroup. When $M$ is not compact the
situation becomes even more complex as one will see later on. This
ends the mathematical presentation of the problem.
\\
Although the problem is purely mathematical - and it will also be
treated in this way - defining a fully diffeomorphism invariant
topology on $\mathbb{L}(M) $ is an important issue in physics,
especially in general relativity and quantum gravity.  In general
relativity and more accurately in cosmology, one would like to
make a gauge invariant perturbation of a given space time (for
example a Robertson - Walker space time).  The constructions in
~\cite{Bombelli,Bombelli2,Bombelli3} propose an answer to this
problem.  The solution in ~\cite{Bombelli,Bombelli2} however has a
few problems.  The defining functions of the uniform topology
generated by the pseudodistance $\bar{d}$ on $ \mathbb{L} ( M ) $
are too sensitive to small perturbations of future and past
distinguishing metrics. As an illustration of this unfortunate
property, the authors show that a (non conformal) perturbation of
compact support $\delta g$ of the Minkowski metric $ \eta $ on $
\mathbb{R}^{d+1} $ is at maximal distance 1 from the latter. This
means that the mappings $ \psi \rightarrow \bar{d}( g ,
\psi^{*}\tilde{g} )$ ($g \neq \tilde{g} $ future and past
distinguishing) are not continuous at the identity in $Diff(M)$
equipped with the Schwartz topology. This is a property which I
think is in contrast with our physical intuition.  In this article
we modify the definitions of ~\cite{Bombelli,Bombelli2} in such a
way that - as mentioned before - the actions $f_{g, \tilde{g} }$
are continuous - at least when $g$ and $\tilde{g}$ are globally
hyperbolic.  Since we do not give up diagonal diffeomophism
invariance (at least in the compact case), we could at this point
do exactly the same as in ~\cite{Bombelli,Bombelli2} by taking the
quotient $\mathbb{L} (M)/Diff(M) $ and hence obtaining a fully
diffeomorphism invariant topology on $ \mathbb{L} (M) $.  Taking
the quotient however might lead us to considerable problems
concerning the Hausdorff property of the quotient topology.  We
will touch upon this very delicate issue in the epilogue. The
problem is that $ Diff(M) $ is just a too big object to handle!
Our proposal - as mentioned before - is to give up the idea of a
fully diffeomorphism invariant topology, and reduce the invariance
to a maximal amenable subgroup. In this way we hope to recover a
finer topology.  We remark that in our construction as well as in
~\cite{Bombelli,Bombelli2} the manifold $M$ is kept fixed.  On the
other hand is a fully diffeomorphism invariant metric topology on
$\mathbb{L}(M) $ is a vital ingredient in some formulations of
quantum gravity where one wants to take a "sum" over isometrically
or topologically inequivalent (compact) space times. In fact such
a program has only been rigorously performed in two space time
dimensions, where the genus of a compact orientable surface
provides the parameter which distinguishes between homeomorphism (
= diffeomorphism) classes of surfaces.  In dimensions 3 and 4 one
does not have a classification of homeomorphism equivalent
manifolds yet.  The problem becomes even more difficult in
dimension 4 in the sense that topological structure is not the
same as differential structure; Donaldson gave examples of four
dimensional topological spaces which do not have a smooth
structure, he also constructed four dimensional topological spaces
which have an infinite number of inequivalent smooth structures
~\cite{Donaldson}. In Euclidian quantum gravity, the most serious
attempt until now is the dynamical triangulations approach which
is inspired on results of Gromov, who proved that the space of
isometry classes of compact metric spaces is complete in the
Gromov-Hausdorff topology.  This approach is investigated by J.
Ambj\o rn, R. Loll, M. Carfora and others ~\cite{Carfora}. They
have proven that every Riemannian structure $(M,h)$ of bounded
geometry is the Gromov Hausdorff limit of a sequence of
dynamically triangulated manifolds (this is a piecewise linear
manifold with fixed edge length) - so in this formalism one takes
the sum over inequivalent isometry structures. This approach
however, makes heavy use of the properties of a Riemannian metric.
Bombelli defined in ~\cite{Bombelli3} a distance on isometry
classes of Lorentzian structures by introducing the original idea
of statistical Lorentzian geometry and as such tried to do the
same for Lorentzian structures as Gromov did for the Riemannian
counterpart.  This construction however is restricted to
Lorentzian structures $(M,g)$ such that $M$ has finite volume with
respect to the invariant volume form defined by $g$; we believe
that a generalization of the results in ~\cite{Bombelli3} to
structures of infinite volume is very unlikely without braking
isometry invariance.  We think however that further investigation
of ~\cite{Bombelli3} is needed.  In the epilogue of this article,
we will propose a generalization a la Gromov, which defines a
pseudodstance on the space of isometry classes of compact, future
and past distinguishing structures.

\section{Definitions and an example}
From now on the reader may assume that all Lorentz metrics are
globally hyperbolic, however this severe restriction is absolutely
not necessary and the interested reader may find in appendix A the
definition of "Class $ \mathcal{A} $" space times which is more
adapted to our needs.  Before we state our main definitions we
explain some notations:
\begin{itemize}
\item The symmetric difference $ A \bigtriangleup B $ of two sets
$A$ and $B$ is defined as:
$$ A \bigtriangleup B = (A \setminus B) \cup (B \setminus A)$$
\item The Alexandrov sets $A(p,q)$ are defined as the set of all
points $s$ which are in the causal future of $p$ and the causal
past of $q$.
\end{itemize}
Since all metrics are assumed to be globally hyperbolic the
Alexandrov sets are compact; however the Alexandrov sets for
"Class $ \mathcal{A} $" space times are not necessarily closed.
We will modify now the definitions in ~\cite{Bombelli, Bombelli2}
in an appropriate way:
\\
\newtheorem{deffie}{definition}
\begin{deffie}
With $ g $ and $ \tilde{g} \in \mathbb{L} (M)$ and $ \forall p, q
\in M $ $ A(p,q)$ and $ \tilde{A} (p,q) $ the Alexandrov sets for
$g$ respectively $ \tilde{g} $, define:
\begin{enumerate}
\item  $$ \alpha_{g \tilde{g}}(p,q) =  \begin{cases}
\frac{ V( A(p,q) \bigtriangleup \tilde{A}(p,q)) }{ V( A(p,q) \cup \tilde{A}(p,q)) } & \text{if} \quad 0 < V( A(p,q) \cup \tilde{A}(p,q)) < \infty  \\
 0 & \text{otherwise}
\end{cases} $$
\item  Put $W_{i} \subset W_{i+1} \subset M$, $\cup_{ i \in \mathbb{N}}W_{i} = M$ and the closure of $ W_{i} $ is compact.  Put
$f : {\mathbb{R}}^{+} \rightarrow {\mathbb{R}}^{+}$ an increasing
function such that $\exists \sigma \geq 1 : \quad f(ax) \leq
a^{\sigma}f(x) \quad \forall a \geq 1 $ and $ x \in \mathbb{R}^{+}
$.  Define then the following set of functions: $$ d^{i}_{
\text{cau}} (g, \tilde{g}) = \frac{1}{ f( V( W_{i} )) } \int_{
W_{i} \times W_{i}} { \alpha_{g, \tilde{g}}(p,q) dV(p) dV(q) }$$
\item $$d_{\text{vol}}^{i} (g , \tilde{g} ) = \sup_{p \in W_{i}} \mid ln \left( \frac{ \sqrt{ - \mid g(p) \mid } }{ \sqrt{ - \mid \tilde{g}(p) \mid } } \right) \mid $$
so this is a pseudodistance which measures the difference in
volume elements.
\item $$ d_{\text{geo}}^{i} ( g , \tilde{g} ) = \sup_{p,q \in W_{i} } |
\lambda(p,q) - \tilde{\lambda} ( p,q) | $$ where $ \lambda (p,q) $
is zero when $ q \notin J^{+} ( p ) $ and otherwise it equals $
\sup_{ \gamma \in C(p,q) } L[ \gamma ]$.  We refer the reader for
the definition of $L$ to ~\cite{Hawking1}.
 When $g$ is globally hyperbolic it has been proven ~\cite{Hawking1} that $ \lambda $
is continuous in $p$ and $q$, moreover one has that $\lambda(p,q)$
equals the length of a non - spacelike $g$ - geodesic curve from
$p$ to $q$.  $ d_{\text{geo}}^{i} ( g , \tilde{g} )$ is clearly a
pseudodistance.
\end{enumerate}
\end{deffie}
In the introduction we already stated the main advantage of $ d^{
i}_{\text{cau}} $ over the supremum definition:
$$ \sup_{p,q \in M:
V(A(p,q) \cup \tilde{A}(p,q)) \geq \zeta } \alpha_{g
\tilde{g}}(p,q)
$$ where $ \zeta $ is a minimal volume scale that has been
introduced to avoid that points which are "too close" and are
causally related in $g$ ($\tilde{g}$) but not in $\tilde{g}$ ($g$)
put $\alpha_{g \tilde{g}}$ ($\alpha_{\tilde{g} g})$ and hence the
supremum equal to $1$ ~\cite{Bombelli}. But this is not the only
problem of the definition, there exist also cases where the points
p and q are "far away" and define for one metric a very tiny
Alexandrov set of considerable volume and of zero volume for the
other one. The function $ d^{ i}_{\text{cau}} $ distinguishes
between conformally inequivalent strongly causal metrics (see
proposition 3) as does the supremum definition for future and past
distinguishing metrics.  The pseudodistance $d_{\text{vol}}^{i} $
compares the volume forms defined by both Lorentz metrics.  We
will use it to "symmetrize" the distance functions $d^{i}_{
\text{cau}}$ . The last metric compares "the geodesic length
between the points p and q", this tells us something about the
shape of the Alexandrov sets and thus delivers extra information.
The author wishes to stress that for future and past
distinguishing Lorentz metrics $g$ and $\tilde{g}$, $d^{i}_{
\text{geo}}(g, \tilde{g}) > 0$ if and only if $g$ and $\tilde{g}$
are conformally inequivalent.  This makes our topology more
restrictive, which is necessary since the definition of
$d^{i}_{\text{cau}}$ is a considerable relaxation of the supremum
definition.  One could give an other definition of $d^{i}_{
\text{geo}}$ by introducing another parameter $ \epsilon $ as
follows:
$$d_{\epsilon}^{i} ( g ,\tilde{g} ) = \sup_{p,q \in W_{i} : V(A(p,q)), V(\tilde{A}(p,q)) \geq \epsilon} \mid
ln \left( \frac{ \lambda(p,q) }{\tilde{\lambda}(p,q)} \right) \mid
$$  We do not choose this distance function because the function $
\epsilon \rightarrow d_{ \epsilon}^{i} ( g, \tilde{g} ) $ is not
upper continuous when $g, \tilde{g} $ are globally hyperbolic and
$M$ is compact.  We will come back to this later on.  We will
discuss the use of the function $f$ after proposition 2.
\\ The reader might ask if the resulting topology does
depend on the choice of the sequence $W_{i}$: we will prove this
is not the case. In the sequel we will follow quite accurately the
structure of ~\cite{Bombelli}, but we make some crucial
modifications where necessary. We call two metrics $g$ and $
\tilde{g} \quad ( i , \zeta , \epsilon , \alpha ) $ close if and
only if
$$
\begin{cases}
d^{i}_{\text{vol}}(g , \tilde{g}) \leq \zeta \\
d^{i}_{ \text{cau}} ( g ,\tilde{g} ) \leq \epsilon \\
d^{i}_{ \text{geo} } (g , \tilde{g} ) \leq \alpha \end{cases} $$
It is straightforward that if $g$ and $ \tilde{g} $ are\ $ ( i ,
\zeta , \epsilon , \alpha ) $ close that they are $ (  j, \zeta ,
\frac{ f( V(W_{i}) )}{ f( V( W_{j}) )} \epsilon, \alpha  ) $ close
$ \forall j \leq i $.  In the two following propositions we prove
a generalised symmetry and transitivity property.
\newtheorem{theo}{Proposition}
\begin{theo}
If $(g , \tilde{g})$ are $ ( i , \zeta , \epsilon , \alpha  ) $
close then $( \tilde{g} , g )$ are $ ( i , \zeta ,  e^{( 4 +
\sigma ) \zeta } \epsilon , \alpha ) $ close.
\end{theo}
\emph{Proof} \\ $ d^{i}_{\text{vol}}( \tilde{g}, g ) \leq \zeta $
is obvious because $ d^{i}_{\text{vol}} $ is a pseudodistance. The
inequality for $d^{i}_{ \text{cau}}$ follows from the fact that
for all Lebesgue measurable regions $\mathcal{O}$ :
$$ e^{ -\zeta} \tilde{V} (O) \leq V( \mathcal{O}) \leq e^{ \zeta } \tilde{V} (\mathcal{O}) $$
The Alexandrov sets are Lebesgue measurable ~\cite{Szabados} and
consequently:
$$ \frac{ \tilde{V}( A(p,q) \bigtriangleup \tilde{A}(p,q)) }{ \tilde{V}( A(p,q) \cup \tilde{A}(p,q)) } \leq e^{ 2 \zeta }
\frac{ V( A(p,q) \bigtriangleup \tilde{A}(p,q)) }{ V( A(p,q) \cup
\tilde{A}(p,q)) } $$ for all $p$ and $q$ such that $ V(A(p,q) \cup
\tilde{A}(p,q)) > 0$. Using this property we get that:
$$ \frac{1}{ f( \tilde{V}( W_{i} )) } \int_{ W_{i} \times W_{i}}
{ \alpha_{\tilde{g}, g}(p,q) d\tilde{V}(p) d\tilde{V}(q) } \leq
\frac{f(V( W_{i} ))}{ f( \tilde{V}( W_{i} )) } e^{ 4 \zeta }
d^{i}( g, \tilde{g} )$$  Because $ V(W_{i}) \leq e^{ \zeta}
\tilde{V} (W_{i}) $, the properties of $f$ imply that: $$
\frac{f(V( W_{i} ))}{ f( \tilde{V}( W_{i} )) } \leq e^{ \sigma
\zeta} $$ which gives the result. \\ That $d^{i}_{ \text{geo} }
(\tilde{g}, g) \leq \alpha$ is obvious. $\square$
\\
The proof clearly shows why the properties of $f$ were necessary
if we later want to define a uniformity.  One also immediately
remarks that $ \forall  j , \gamma , \delta , \nu $ there exist $
\ i , \zeta , \epsilon , \alpha  $ such that if $ (g, \tilde{g} )$
are $ (  i , \zeta , \epsilon , \alpha  )$ close then $ (
\tilde{g} ,g) $ are $ ( j , \gamma , \delta , \nu  )$ close.
\\
\begin{theo}
If $ (g, \tilde{g}) $ and $ (\tilde{g}, \bar{g} ) $ are $( i ,
\zeta,  \epsilon , \alpha  ) $ close then $ (g,\bar{g}) $ are $ (
 i , 2  \zeta ,  2(1 + e^{ ( 4 + \sigma ) \zeta } ) \epsilon , 2 \alpha  )
$ close.
\end{theo}
\emph{Proof} That $ d^{i}_{\text{vol}} ( g,\bar{g}) \leq 2 \zeta $
is obvious because $ d^{i}_{\text{vol}}$ is a pseudodistance. To
prove the second assertion assume $p$ and $q$ are points such that
$$ \alpha_{g \tilde{g}} (p,q) , \alpha_{ \tilde{g} \bar{g} }(p,q)
< 1$$ We start by observing the following inequality:
\begin{eqnarray*} V( A(p,q) \bigtriangleup \bar{A} (p,q) ) & \leq
&  V( A(p,q) \bigtriangleup \tilde{A} (p,q) )+  V( \tilde{A}(p,q)
\bigtriangleup \bar{A} (p,q) ) \\ & \leq & V( A(p,q)
\bigtriangleup \tilde{A} (p,q) )+  e^{ \zeta } \tilde{V}(
\tilde{A}(p,q) \bigtriangleup \bar{A} (p,q) )
\end{eqnarray*}
so that:

$$ \frac{V( A(p,q) \bigtriangleup \bar{A} (p,q) )}{ V(A(p,q) \cup \bar{A} (p,q)
)} \leq \frac{V( A(p,q) \bigtriangleup \tilde{A} (p,q) )}{
V(A(p,q) \cup \bar{A} (p,q) )} + e^{ \zeta } \frac{V(
\tilde{A}(p,q) \bigtriangleup \bar{A} (p,q) )}{ V(A(p,q) \cup
\bar{A} (p,q) )} $$ We can estimate that:  \begin{eqnarray*} V(
A(p,q) \cup \tilde{A} (p,q) ) & \leq & V( A(p,q) \cup \bar{A}
(p,q) ) + V( A(p,q) \bigtriangleup \tilde{A} (p,q) ) \\ & \leq &
V( A(p,q) \cup \bar{A} (p,q) ) + \alpha_{g \tilde{g}} (p,q)V(
A(p,q) \cup \tilde{A} (p,q) )
\end{eqnarray*}
So we have that:
$$ V( A(p,q) \cup \bar{A} (p,q) ) \geq ( 1 - \alpha_{g \tilde{g}}
) V( A(p,q) \cup \tilde{A} (p,q) ) $$ A similar calculation shows
that:
$$ \alpha_{g \bar{g}}(p,q) \leq \frac{\alpha_{g \tilde{g}}(p,q)}{1 - \alpha_{g
\tilde{g}}(p,q)} + e^{ 2 \zeta } \frac{\alpha_{\tilde{g}
\bar{g}}(p,q)}{1 - \alpha_{\tilde{g} \bar{g}}(p,q)} $$ Now the
function $ x \rightarrow \frac{x}{1 -x } $ is monotonically
increasing and exceeds $1$ at $ x = \frac{1}{2} $. Because the
function $\alpha_{g \bar{g}}(p,q)$ can at most obtain the value
$1$, we can conclude that:
$$ \alpha_{g \bar{g}}(p,q) \leq 2 \alpha_{g \tilde{g}}(p,q) + 2 e^{ 2 \zeta } \alpha_{\tilde{g}
\bar{g}}(p,q) $$ for all $p$ and $q$.  Using this result, we
obtain that :
$$ d^{i}_{ \text{cau}}( g, \bar{g} ) \leq 2 d^{i}_{ \text{cau}} ( g , \tilde{g} ) + 2 e^{ ( 4 + \sigma )
\zeta } d^{i}_{ \text{cau}} ( \tilde{g} , \bar{g} ) $$ which
proves the claim. $\square$
\\ Again it is easy to see that
$ \forall  j , \gamma , \delta , \nu  $ there exist $  i , \zeta ,
\epsilon , \alpha  $ such that if $ (g, \tilde{g} )$ and $
(\tilde{g} , \bar{g} ) $ are $ (  i , \zeta , \epsilon , \alpha )$
close then $ ( g ,\bar{g}) $ are $ ( j , \gamma , \delta , \nu )$
close.
\\
\newtheorem{rem}{remark}
\begin{rem}
All the previous properties remain unchanged when we replace $
\alpha_{g \tilde{g}}(p,q) $ by $ z^{ \beta }( \alpha_{g
\tilde{g}}(p,q) ) $ where $ \beta \in {\mathbb{R}}^{+} $ , $z$ is
a $C^{2}$ convex function with $ z(0) = 0 $ and $ \frac{dz}{dx}(0)
\geq 0 $ and $ z^{ \beta } (x) = z ( \beta  x) $ . This is the
consequence of the next inequality: $$  z^{ \beta } ( \alpha_{g
\bar{g}}(p,q) ) \leq \frac{1}{e^{ 2 \zeta } + 1 } z^{2 \beta ( 1 +
e^{ 2 \zeta })} ( \alpha_{g \tilde{g}}(p,q))  + \frac{e^{ 2 \zeta
}}{e^{ 2 \zeta } + 1 } z^{2 \beta ( 1 + e^{ 2 \zeta })} (
\alpha_{\tilde{g}\bar{g}}(p,q)) $$ One also immediately notices
that the resulting topology will remain unchanged.
\end{rem}
Now we will say more about the function $f$ introduced in
definition 1.  Let us introduce the following notations:
\begin{itemize}
\item $d'_{\text{cau }} ( g , \tilde{g} ) = \limsup_{ i \to \infty
} d^{i}_{\text{cau }}$
\item $d_{ \text{vol }}(g, \tilde{g}) = \sup_{p,q \in M} \mid ln \left(
 \frac{ \sqrt{ - \mid g(p) \mid } }{ \sqrt{ - \mid \tilde{g}(p) \mid } } \right) \mid $
\item $ d_{\text{geo}} ( g , \tilde{g} ) = \sup_{p,q \in M } |
\lambda(p,q) - \tilde{\lambda} ( p,q) | $
\end{itemize}
and as before we define $g$ and $\tilde{g}$ to be $( \zeta,
\epsilon , \alpha ) $ close if :
$$
\begin{cases}
d_{\text{vol}}(g , \tilde{g}) \leq \zeta \\
d'_{\text{cau}} ( g ,\tilde{g} ) \leq \epsilon \\
d_{\text{geo} } (g , \tilde{g} ) \leq \alpha \end{cases} $$ The
reader can easily check then that the following modified versions
of proposition 1 and 2 are valid:
\begin{itemize}
\item if $(g , \tilde{g})$ are $ (  \zeta , \epsilon , \alpha  ) $
close then $( \tilde{g} , g )$ are $ ( \zeta ,  e^{( 4 + \sigma )
\zeta } \epsilon , \alpha ) $ close
\item if $ (g, \tilde{g}) $ and $ (\tilde{g}, \bar{g} ) $ are $ (\zeta,  \epsilon , \alpha  ) $
close then $ (g,\bar{g}) $ are $ ( 2  \zeta ,  2(1 + e^{ ( 4 +
\sigma ) \zeta } ) \epsilon , 2 \alpha  ) $ close
\end{itemize}
One can also give appropriate versions of the remarks concerning
propositions 1 and 2.  This shows - as the reader will understand
later - that the functions $d_{\text{vol}} , d'_{\text{cau}}, d_{
\text{geo} }$ define a uniform topology. There are however two
objections against this construction:
\begin{itemize}
\item The function $d'_{\text{cau }} $ depends on the sequence
$W_{i}$ and is therefore not canonically determined by $M$.
\item  This topology depends on the function $f$ for which there doesn't seem to be a good
proposal.  The following example will force us to make such a
proposal for $f$, and there is no guarantee that this is a good
choice under all circomstances.
\end{itemize}
\newtheorem{exie}{example}
\begin{exie}
We consider flat two dimensional Minkowski space - time with the
usual metric $ \eta = \left( \begin{array}{cc}
-1 & 0  \\
0 & 1 \end{array} \right) $.  Now we make a perturbation $ \delta
g$ of this metric of compact support $S$.  For computational
simplicity we will assume that $S$ is the rectangle of length m
and height l centered around the origin. The perturbed metric
looks like: $$ \eta + \delta g = \eta  + \epsilon \chi_{S} (1,-1)
\otimes (1, -1)
$$ where $ \epsilon
> 0$ is small and $ \chi_{S} $ is the characteristic function of
support $S$. Furthermore we define sets $ W_{s} = B(0, s) \quad s
\in \mathbb{N}_{0}$ and assume that $ s \gg l,m $. \end{exie}
Using the convention that
 $\alpha \equiv \alpha (\epsilon)$ is the difference in opening angle of the light cones, we get the
 following picture: \\
 \begin{figure}[h]
\postscript{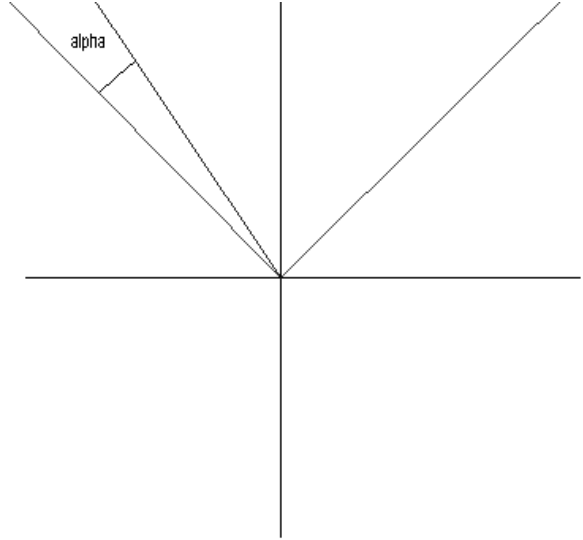}{0.7} \caption{shows the difference in
conformal structure in the perturbed area} \end{figure} We
calculate now $ d_{\text{cau}}^{s} ( g , \tilde{g} )$ for $s
\rightarrow \infty $, to do this we split the double integral in a
few parts. In the sequel the shaded area indicates the range of
the point $q$ we integrate over.  The terminology "maximal order"
indicates the order of the leading term in the radius $s$ of the
domain.  The next picture describes a part of the integral which
contributes a term of maximal order $ \mathcal{O}( s^3 )$ to the
integral.
 \begin{figure}[h] \postscript{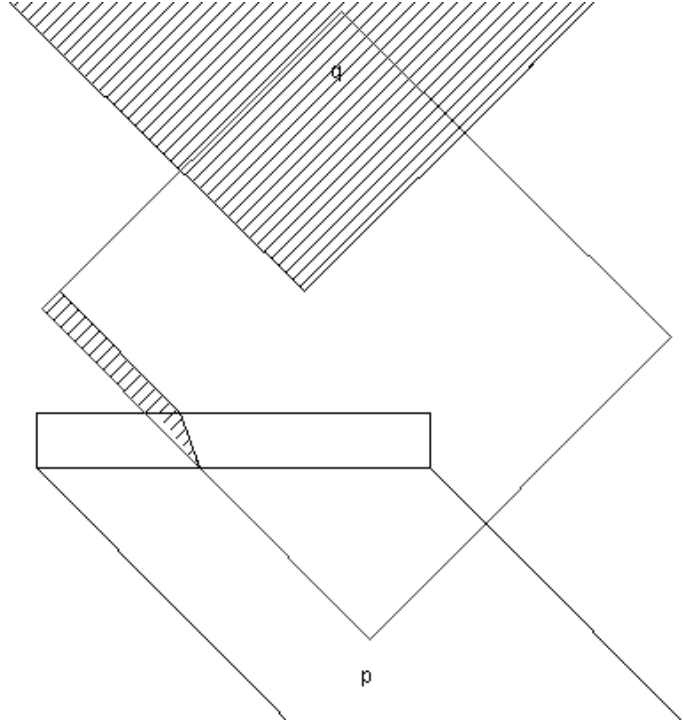}{0.7}
 \caption{gives the picture where q is any point in the
 shaded area and p lies under the rectangle between the parallel
 lines} \end{figure}

\begin{figure}[h] \postscript{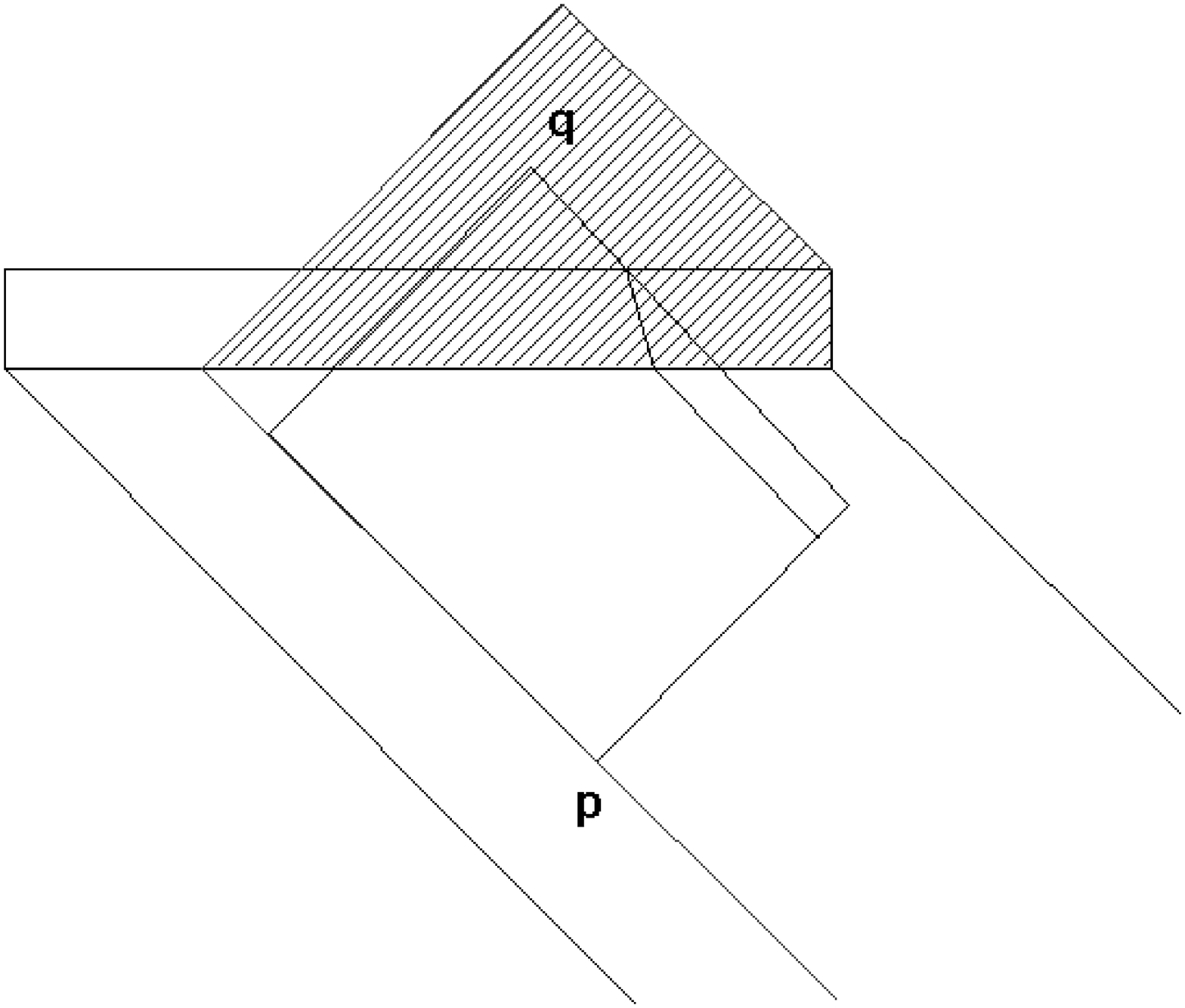}{0.7} \caption{ is of maximal order $ \mathcal{O}  ( s )$} \end{figure}

\begin{figure}[h] \postscript{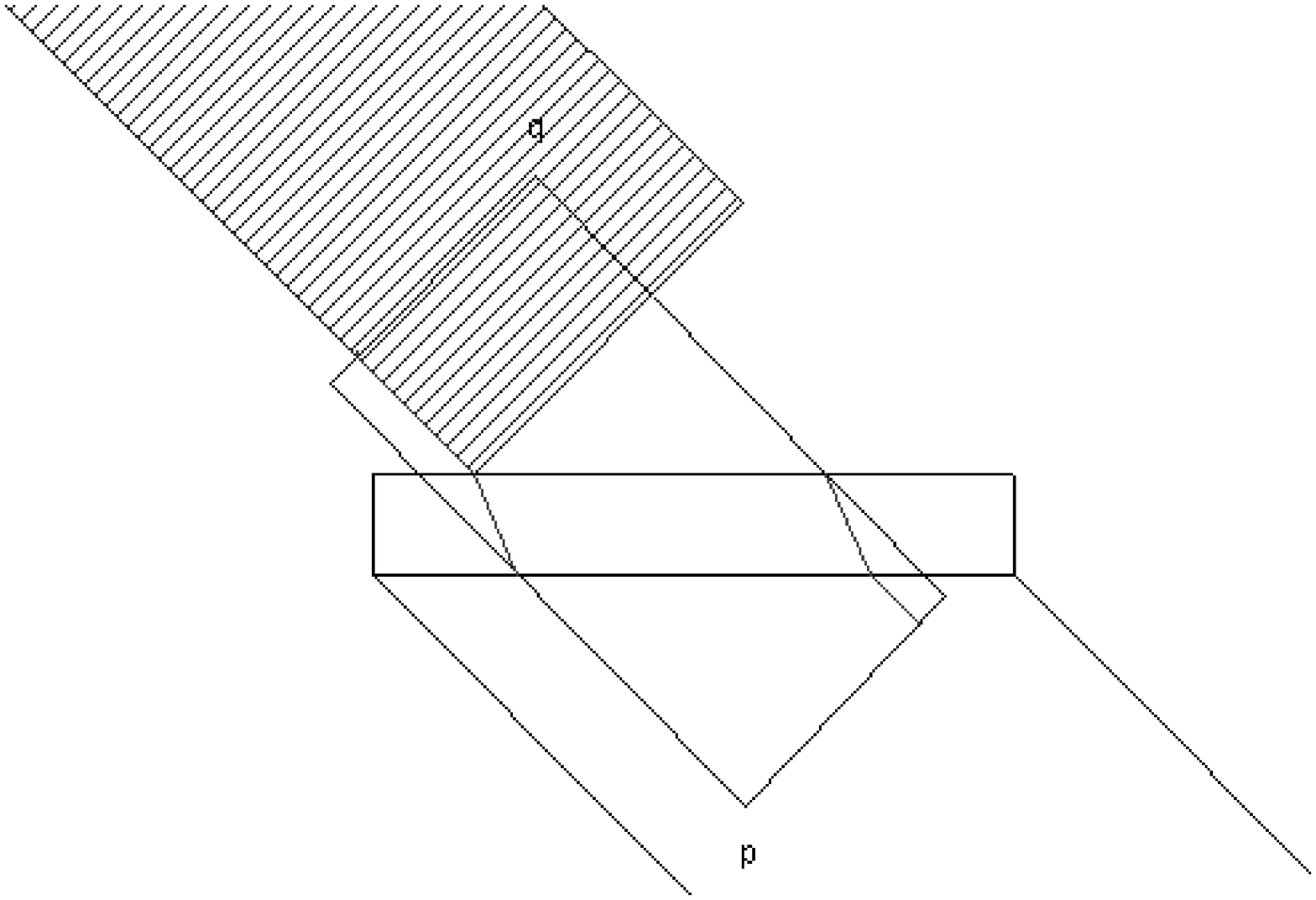}{0.7} \caption{is of maximal order $ \mathcal{O} ( s^2 )$} \end{figure}

\begin{figure}[h]
\postscript{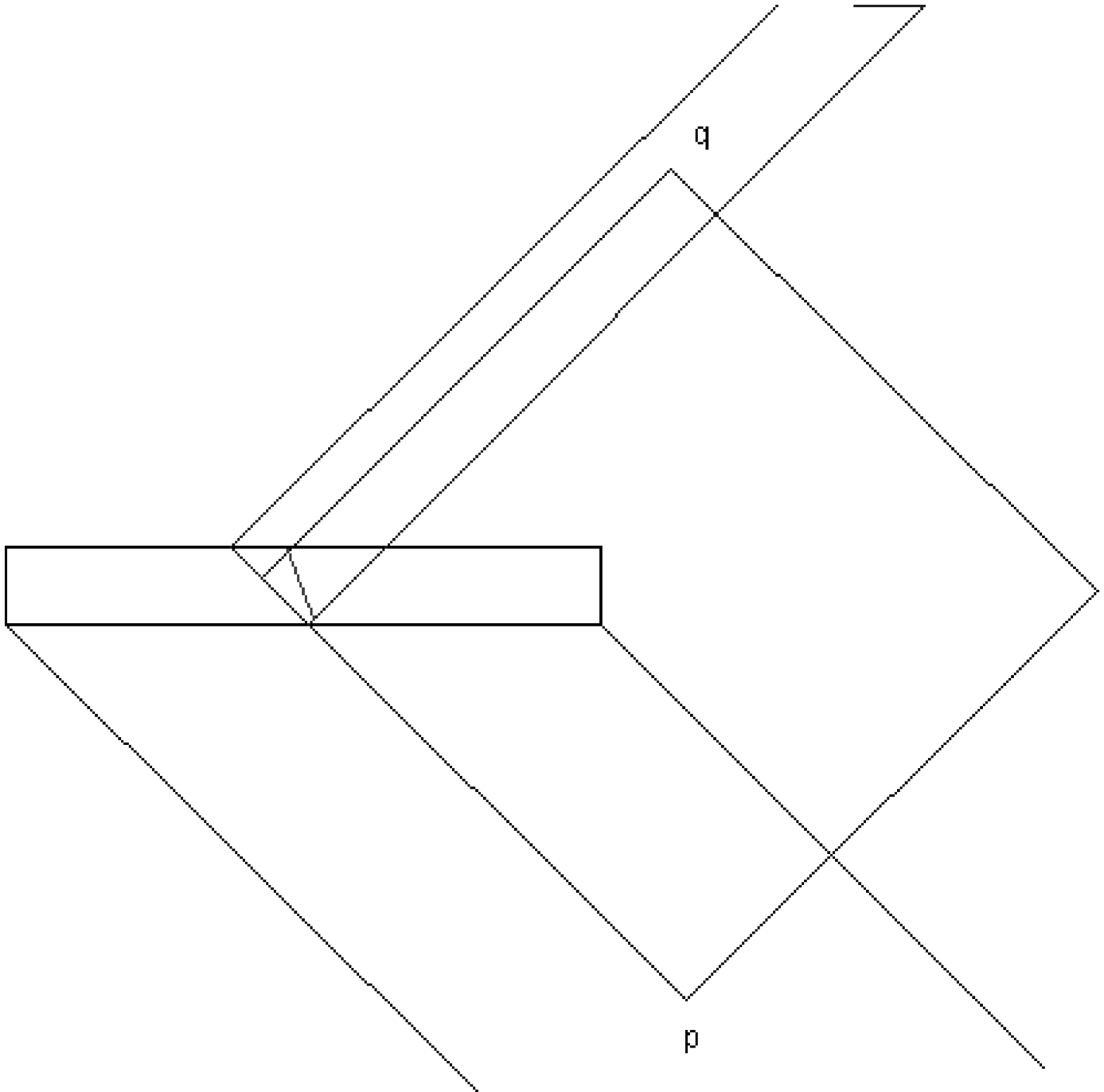}{0.7} \caption{is of maximal order $
\mathcal{O}  ( s^2 )$ } \end{figure}

\begin{figure}[h]
\postscript{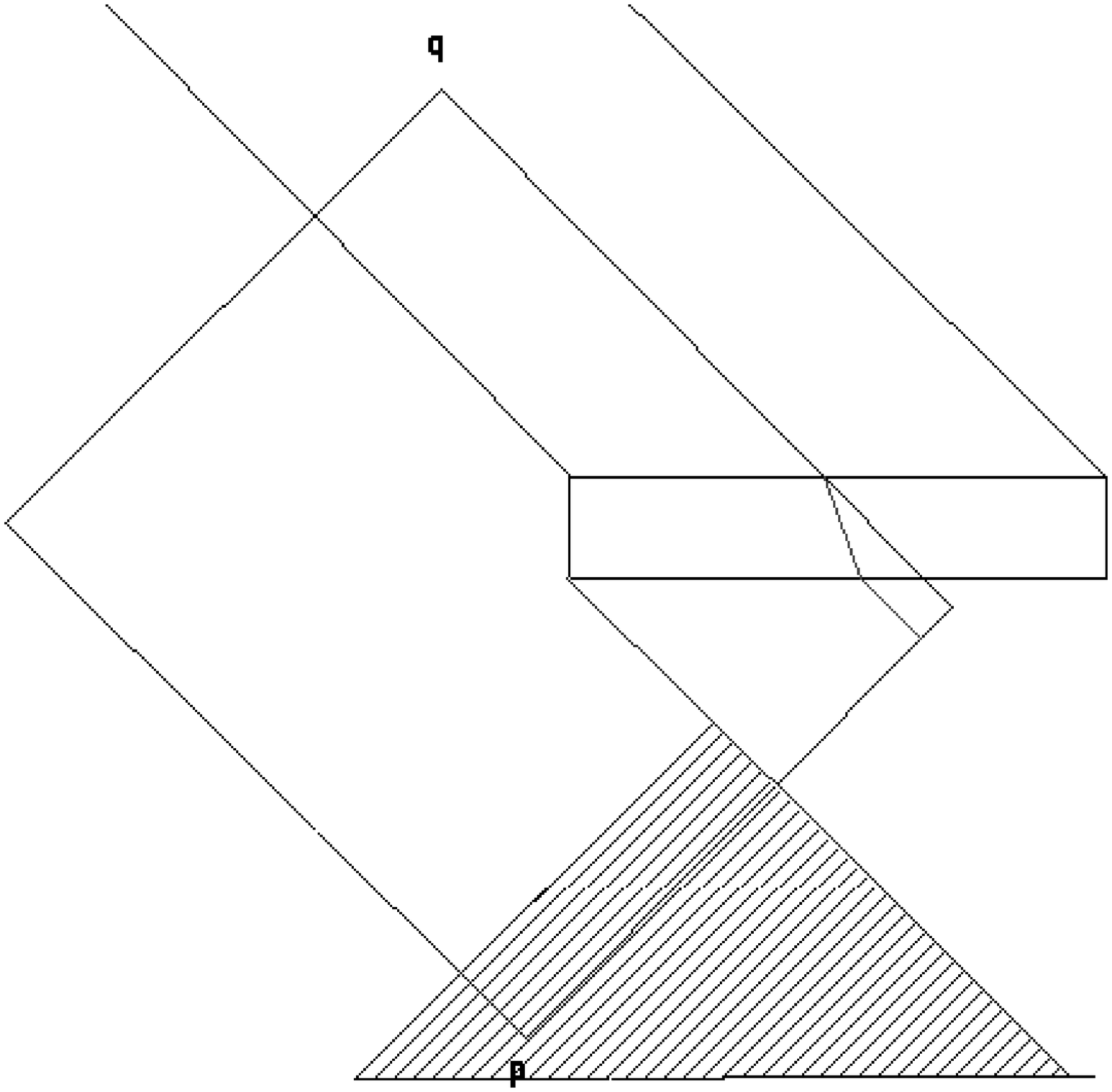}{0.7} \caption{is the symmetric situation of
figure $2$ and contributes at most $ \mathcal{O} ( s^3 )$}
\end{figure}
All other possibilities except the one in figure 6 contribute at
most a factor of $ \mathcal{O}  ( s^2 )$. So if we want to know
the leading order, we should first calculate the contribution
described in figure $2$ . Keep $p$ fixed and put $x,y$ the
lightconecoordinates, we obtain that:
 $$ V( A(p,q) \bigtriangleup \tilde{A}(p,q) ) = l^2 \beta + l^2 \beta^2 + ( x - \delta (p) ) \sqrt{2} l \beta $$
 with $$ \beta = \frac{ tan ( \alpha ) }{ 1 + tan ( \alpha) } $$
and $ \delta(p)$ is the riemannian length of the section of the
null curve for $g$ with initial point $p$ and end point the
intersection of this curve with the top horizontal line of the
rectangle.  One calculates that :
 $$ \int_{W_{i}} \alpha_{g \tilde{g}} (p,q) dV(q) = \int_{ \delta (p)}^{s} dx \int_{ \sqrt{2} l \beta }^{s} dy
 \frac{ l^2 \beta  + l^2 \beta^2 + ( x - \delta (p) ) \sqrt(2) l \beta}{xy} + \mathcal{O} ( s^2) $$
Integration yields:
 $$ \delta (p) \sqrt{2} l \beta ( l^2 \beta + l^2 \beta^2 ) ln \left( \frac{s}{ \delta (p) } \right) ln \left( \frac{s}{ \sqrt{2} l \beta } \right)
 +  \sqrt{2} l \beta ( s - \delta (p) ) ln \left( \frac{s}{ \sqrt{2} l \beta } \right) + \mathcal{O} ( s^2) $$
 Integration over $p$, multiplying by a factor 2 and dividing by
 $f$ gives the following result:
 \begin{eqnarray*} d^{s}_{\text{cau}} ( g , \tilde{g} ) & =
 & \frac{1}{f(V(B(0,s)))} \left( \frac{l \beta m}{\sqrt{2}} s^2 ln ( s ) + \mathcal{O} ( s^2) \right) \\
 & = & \frac{1}{f(V(B(0,s)))} \left( \frac{\beta V(S)}{ \sqrt{2}} s^2 ln ( s ) + \mathcal{O} (
 s^2)\right)
 \end{eqnarray*}
This shows that in $1+1$ dimensions and for $ V(B(0,s)) = x \gg
V(S) $ large enough $d^{s}_{\text{cau }}( \eta , \eta + \delta g )
$ looks like :
$$ d^{s}_{\text{cau }}( \eta , \eta + \delta g ) \sim  \frac{
\beta V(S) }{ 2 \sqrt{2} \pi f(x) } x ln(x) $$ In order to make
the limit $ s \rightarrow \infty $ nonzero and finite, $ f $ has
to be proportional to $ x ln ( x )$ for $ x $ big enough.  It is
easy to see that $ \forall x \geq e $ and $ a \geq 1$ : $$ (ax)
ln(ax) \leq a^2 x ln(x) $$  Therefore it is convenient to define:
$$ f(x) =
\begin{cases}
x & \text{if} \quad 0 \leq x \leq e  \\
 x log(x) & \text{otherwise}
\end{cases} $$
This shows -as anticipated before- that the asymptotics of $f$ is
completely determined by this example modulo a proportionality
constant.  The author tried to check if one gets the same result
in higher dimensions, but this seemed to be an impossible task
(for example the calculation of the intersection of regular
volumes is not a triviality). Probably the only approach here is
the numerical one.  This example also shows that the result
depends on the volume of the distorted area and a "deflection
parameter" $\beta$ which is obviously a much better result than
the one obtained in ~\cite{Bombelli,Bombelli2}.
\\
All the previous remarks lead us to the conclusion that in the non
compact case we have to make another approach than just taking the
limsup and therefore we have to dispose of $d'_{\text{cau}}$.  In
section 6 the reader will see that our final proposal for a
topology does not depend of the function $f$ and therefore the
reader can assume that $f=1$ unless stated otherwhise.
\section{Properties of the distance functions when certain causality requirements are satisfied}
We begin this section by a review some terminology
~\cite{Hawking1}.
\\
\textbf{terminology}
\\
\begin{itemize}
\item $ p \prec q $ if and only if $ \exists $ a future oriented timelike
curve from $ p $  to $ q $
\item $ p \preceq q $ if and only if $ \exists $ causal
curve from $ p $ to $ q $
\item $ p \rightarrow q $  if and only if $ q \in J^{+}
(p) \backslash I^{+} ( p ) $
\item If O is open in M,  $ \prec_{O} $  , $ \preceq_{O} $
and $ \rightarrow_{O} $ are defined analogously with the extra
requirement that all connecting causal ( timelike ) curves must
lie in $O$.  These are the partial order relations defined by the
sets $ I^{\pm} ( p,O) , J^{\pm} ( p , O ) $
\item The \textbf{restriction} of $ \prec $  to O  $ \prec_{ | O } $ is
defined in the usual way. \item Define by $C(p,q)$ the set of all
$C^{0}$ causal curves with initial end point $p$ and final
endpoint $q$.
\item A map $f$ from a topological space $O$ to the set of all subsets of $O$
is called outer continuous in a point $p \in O$ if and only if $
\forall K \subset  O \backslash \bar{f(p)}  $ compact, there
exists a neighborhood $V$ of $p$ such that for all $ r \in V $ it
follows that $ f(r) \cap K = \emptyset $
\item A space - time $ ( M , g )$ is causally continuous if and only if the map
$$ I^{+} : p \rightarrow I^{+} ( p) $$ is outer continuous.
\end{itemize}
\begin{rem}
In the definition of outer continuity for the map $f$ it is
usually assumed that $ f(p) $ is open $ \forall p \in O$.  We do
not follow this convention here because it is not essential.
\end{rem}
\begin{rem}
On $C(p,q)$ we define a topology with the following basis:
$$ O_{W} ( \gamma ) = \{ \lambda \in C(p,q) | \lambda \subset W \} $$
$W$ is an open neighborhood of $ \gamma $ in $M$.  Now $g$ is
globally hyperbolic on $M$ if and only if $g$ is strongly causal
and $C(p,q)$ is compact $\forall p,q \in M$  ~\cite{Hawking1} .
This implies that if $K_{1}, K_{2} $ are compact then $ C(K_{1},
K_{2}) $ is compact. Another (equivalent) criterion is that $g$ is
strongly causal and $ A(p,q)$ is compact $ \forall p,q \in M$.
\end{rem}
With these concepts, we can now prove the following theorem.

\begin{theo}
If $ g $ is future and past distinghuishing and $ \tilde{g} $ is a
strongly causal metric which are not conformally related in $ p
\in M $ then $ \exists i_{0} \in \mathbb{N} $ such that $ \forall
j \geq i_{0} : d^{j}_{ \text{cau }}(g , \tilde{g} ) > 0 $
\\ Proof
\end{theo}
Choose $i_{0}$ large enough such that $p \in W_{i_{0}} $. All the
following sets are assumed to be subsets of $ W_{i_{0}} $. We have
to prove that $ \exists X $ of compact closure such that $ \bar{ X
} \subset I^{+} ( p ) \backslash \tilde{J^{+}} ( p ) $ and an open
subset $ Y $ such that $ \forall r \in Y : X \subset I^{+}(r)
\backslash \tilde{J^{+}} ( r) $. Then $ \alpha_{g \tilde{g}} (
r,s) = 1$ on the productset $ X \times Y $.
\\ Choose $ \tilde{U} $ to be a simple neighborhood of p for $ \tilde{g} $ such that
$\tilde{g}_{| \tilde{U}} $ is causally continuous. Because $
\tilde{g} $ is strongly causal there exists $ \tilde{V} \subset
\tilde{U} $ such that no $ \tilde{g} $ - causal curve intersects $
\tilde{V} $ more than once. This implies that on $ \tilde{V}$, $
\tilde{\prec} $ is equivalent to $ \tilde{ \prec}_{ \tilde{V} } $
and the same for $ \tilde{\preceq} $. Now it is obvious that on $
\tilde{V} \quad \exists X $ of compact closure such that $ \bar{ X
} \subset I^{+} ( p ) \backslash \tilde{J^{+}} ( p ) $ \\
(otherwise $ g $ and $ \tilde{g} $ would be conformally related in
$p$ because both of them are future and past distinghuishing in
$p$). Because of the causal continuity of $\tilde{g}_{| \tilde{U}}
$ there exists a $ Z $ such that $ p \in Z $ and $ \forall r \in Z
: \bar{X} \cap \tilde{J}^{+} ( r, \tilde{V} ) = \emptyset $ but
this implies that $ \bar{X} \cap \tilde{J}^{+} ( r ) = \emptyset $
because of the properties of $ \tilde{V} $. Put $Y = Z \cap I^{-}
( p) $ , this set clearly satisfies all the above properties. $
\square$
\\
This theorem indicates that our construction is not trivial.  We
know that the function $ \alpha_{g \tilde{g}} $ is measurable, we
now prove that it is continuous almost everywhere if $ g $ and $
\tilde{g} $ are both globally hyperbolic.
\begin{theo}
If $ g $ and $ \tilde{g} $ are globally hyperbolic then $
\alpha_{g \tilde{g} } $ is almost everywhere continuous.
\\
proof
\end{theo}
We split the proof in a few parts.
\begin{itemize}
\item Suppose $p,q \in M$ we prove
that the function $ (r,s) \rightarrow A(r,s) $ is outer continuous
in $(p,q)$ (and of course the same works for $ \tilde{g} $).
Assume not, then one has a compact set $ K \subset M \backslash
A(p,q)$, sequences $ r_{n} \rightarrow p $ and $ s_{n} \rightarrow
q$ such that $ r_{n} \prec r_{m} \prec p $ and $ q \prec s_{m}
\prec s_{n} \quad \forall n < m$.  Moreover one has causal curves
$ \lambda_{n} $ with initial endpoint $r_{1}$ and final endpoint
$s_{1} $, passing trough $ r_{n}$ and $ s_{n} $, which intersect
$K$. Because $C(r_{1},s_{1} )$ and $K$ are compact we can find a
subsequence $ \lambda_{n_{k}} \rightarrow \lambda \in
C(r_{1},s_{1} ) $ where $ \lambda $ is a causal curve passing
trough $p$ and $q$ which intersects $K$. This is a contradiction.
\item It is now obvious that the function $ (r,s) \rightarrow A(r,s)
\cup \tilde{A} (r,s) $ is outer continuous in $(p,q)$.  Choose $ K
\subset M \backslash (A(p,q) \cup \tilde{A} (p,q)) $ then $ K
\subset M \backslash A(p,q) , M \backslash \tilde{A} (p,q )$ so
there exist $ R , \tilde{R} $ neighborhoods of $p$ and $ S ,
\tilde{S} $ neighborhoods of $q$ such that $ \forall r \in R \cap
\tilde{R}, s \in S \cap \tilde{S} : K \cap A(r,s) = \emptyset = K
\cap \tilde{A} ( r,s) $ so $ K \cap ( A(p,s) \cup \tilde{A} (r,s)
) =  \emptyset  $.
\item We can now say that $ (r,s) \rightarrow V(A(r,s) \cup
\tilde{A} (r,s)) $ is continuous in $(p,q)$. It is obvious that $
\forall K \subset \overset{ \circ}{A(p,q) \cup \tilde{A} ( p,q)}$
compact $\exists R,S $ neighborhoods of $p$ and $q$ respectively
such that $ \forall r \in R, \quad s \in S : K \subset \overset{
\circ}{A(r,s) \cup \tilde{A} (r,s)} $. Now taken in account that $
A(p,q) \cup \tilde{A} (p,q)$ is compact and $M$ is Hausdorff, $
\forall \epsilon > 0 $ sufficiently small $ \exists K_{\epsilon} $
compact and $U_{ \epsilon } $ open such that $ K_{ \epsilon }
\subset \overset{ \circ}{A(p,q) \cup \tilde{A} (p,q)} \subset
A(p,q) \cup \tilde{A} (p,q)  \subset U_{\epsilon } $ and $ V(U_{
\epsilon }) < V(A(p,q) \cup \tilde{A} (p,q)) + \epsilon $ and $
V(K_{\epsilon})
> V(A(p,q) \cup \tilde{A} (p,q)) - \epsilon $. Because of the
previous remark and the outer continuity of $ (r,s) \rightarrow
A(r,s) \cup \tilde{A} (r,s)$ we can find neighborhoods $R_{
\epsilon}$ of $p$ and $S_{ \epsilon }$ of $q$ such that $K_{
\epsilon } \subset A(r,s) \cup \tilde{A} (r,s) \subset U_{\epsilon
} \quad \forall r \in R_{ \epsilon } $ and $ s \in S_{ \epsilon }$
. The result is now obvious.
\item  With the same techniques it
is easy to prove that $ (p,q) \rightarrow V( A(p,q) \bigtriangleup
\tilde{A}(p,q))  $ is continuous everywhere.
\item The only problem which might arise in the definition of $
\alpha_{g , \tilde{g} } $ is when $ V( A( p,q) \cup \tilde{A} (
p,q ) ) = 0 $. Suppose now that $ A(p,q) = \emptyset =
\tilde{A}(p,q) $ then it is easy to see that exist open
neighborhoods $ X $ of $p$ and $Y$ of $q$ of compact closure such
that $ \forall r \in X, s \in Y : A(r,s) = \emptyset = \tilde{A}
(r,s) $ so the continuity is proven in this case.
\item There is only left the case where $A(p,q) $ or $ \tilde{A}
(p,q) $ is nonempty but has zero volume. This implies that $q \in
\partial J^{+} (p) $ or $q \in \partial \tilde{J}^{+} (p) $, but
this set has zero volume;  the result is now obvious. $\square$
\end{itemize}
In the proof we used the simple fact that $ V(A(p,q)) > 0 $ if and
only if $ p \in I^{-} (q) $ which was proven in ~\cite{Bombelli}.
Proposition $3$ will give the key ingredient for the fact that the
resulting topology on $ \mathbb{L} (M) $ will be Hausdorff on the
subset of strongly causal $C^{2}$ metrics.  It is now interesting
to remark that if $g$ is strongly causal the manifold topology is
the same as the corresponding Alexandrov topology
~\cite{Hawking1}. Moreover the group of conformal $ C^{ \infty }$
diffeomorphisms is in this case equal to the group of the $
\rho_{g} $ - homeomorphisms, where the $ \rho_{g} $ - topology is
defined as the finest topology for which $ E \subset M $ belongs
to $ \rho_{g} $ if and only if $ \forall \gamma $ ($ \gamma$ a $g$
timelike curve) $ \exists O $ open in $ M $ such that $ O \cap
\gamma = E \cap \gamma $.  This fact is surprising because at
first sight the $ \rho_{g} $ - topology is not that much finer
than the manifold topology.  This is a very useful
characterization of the conformal group, so in this case the set
of conformal structures agrees with the quotient of $ \mathbb{L}
(M) $ with the following equivalence relation :
$$ g \sim \tilde{g} \longleftrightarrow \exists \psi \in Diff_{ \rho_{g} } (M)$$
such that $$ \psi^{*} g = \tilde{g} $$ $Diff_{ \rho_{g} } (M)$ is
the subgroup of the $ \rho_{g} $ continuous homeomorphisms
~\cite{Hawking2}.
\\
In the next section we will give a very quick introduction to
uniformities and topologies. The reader who wants a more thorough
treatment is invited to read  ~\cite{Kelley}
\\
\section{ A quick review of uniformities and topologies}
Let $(X,d)$ be a topological space where $d$ is a (pseudo)
distance and denote by $ \tau $ the corresponding locally compact
topology. It is an elementary fact that the open balls $B_{1/n } (
p )$ with radius $1/n : n \in \mathbb{N}_{0} $ around $p$ define a
countable basis for $ \tau $ in $p$.  In this chapter $I,J$ will
denote index sets.  A $(X, \tau)$ cover $C$ is defined as follows:
$$ C = \{ A_{i} | A_{i} \in \tau, i \in I \} $$
such that $$ \bigcup_{ i \in I } A_{i} = X $$ If $C =  \{ A_{i} |
A_{i} \in \tau, i \in I \} , D = \{ B_{j} | B_{j} \in \tau, j \in
J \} $ are $(X, \tau)$ covers then we say that $C$ is finer than
or is a refinement of $D$, $ C < D$ if and only if $$ \forall i
\in I \quad \exists j \in J : A_{i} \subset B_{j} $$  Next we
define a few operations on the set of covers $C(X, \tau )$:
\\
\textbf{Operations on covers}
\begin{itemize}
\item Let $C,D$ be as before,  $$ C \wedge D  = \{ A_{i} \cap B_{j} | A_{i}, B_{j} \in \tau \quad
i \in I, j \in J \}$$ $C \wedge D$ is obviously a cover, moreover
the doublet $ C(X ,\tau),\wedge $ is a commutative semigroup.
\item For $ A \subset X$ the star of $A$ with respect to $C$ is
defined as follows:
$$ St(A,C) = \cup_{ A_{i} \in C: A \cap A_{i} \neq \emptyset}
A_{i} $$
\item The star of $C, C^{*}$ is then defined as:
$$ C^{*} = \{ St(A_{i},C) | A_{i} \in C \} $$
Remark that $ C < C^{*} < C^{**} \ldots $ and that if $I $ is
finite then there exists a $n \in \mathbb{N} $ such that after $n$
star operations $C$ has become the trivial cover.
\end{itemize}
Using the topological basis of open balls, we can define
elementary covers $C_{n} \quad n \in \mathbb{N}_{0} $ as follows:
$$ C_{n} = \{ B_{1/n } ( p ) | p \in X \} $$
These elementary covers now define a subset $U$ of $C(X, \tau )$ :
$$ U = \{ C \in C(X, \tau )| \exists C_{n} : C_{n} < C \} $$ The set $U$
satisfies the following obvious properties:
\begin{enumerate}
\item If $ C \in U $ and $ C < D $ then $ D \in U $
\item If $C,D \in U $ then $ C \wedge D \in U $
\item If $ C \in U $ then $ \exists D \in U : D^{*} < C $
\end{enumerate}
From now on we take the above properties as a \textbf{definition}
for a \textbf{uniformity}:
\begin{deffie}
Let $X$ be a set, a cover $C $ is defined as:
$$ C = \{ A_{i} | A_{i} \subset X, i \in I \} $$
such that $$ \bigcup_{ i \in I } A_{i} = X $$ A collection of
covers $U$ is called a \textbf{uniformity} for $X$ if and only if
\begin{enumerate}
\item If $ C \in U $ and $ C < D $ then $ D \in U $
\item If $C,D \in U $ then $ C \wedge D \in U $
\item If $ C \in U $ then $ \exists D \in U : D^{*} < C $
\end{enumerate}
where all definitions of $ < , \wedge $ and $^{*}$ are independent
of $ \tau $.
\end{deffie}
It has been proven that any uniformity can be generated by a
family of pseudodistances ~\cite{Page}.  This indicates a
uniformity defines a topology.  For our applications we need a
different ingredient.
\begin{deffie}
Let $I$ be a directed net, and suppose $B_{i}(x) \subset X$
satisfy the following properties:
\begin{enumerate}
\item $x \in B_{i} ( x ) \quad \forall x \in X, i \in I $
\item If $ i \leq j $ then $ B_{i} (x) \subset B_{j} (x) \quad
\forall x \in X $
\item $ \forall i \in I , \exists j \in I$  such that $ \forall y \in
B_{j}(x): x \in B_{i} (y)$
\item $ \forall i \in I, \exists j \in I$ such that if $ z \in
B_{j} (y), y \in B_{j}(x) $ then $ z \in B_{i} (x) $.
\end{enumerate}
then we call the family of all $B_{i}(x)$ \textbf{a uniform
neighborhood system}.
\end{deffie}
Now it has been proven that if $ \{ B_{i}(x) | x \in X, i \in I
\}$ is a uniform neighborhood system then the family of covers:
$$ C_{i} = \{ B_{i} (x) | x \in X \} $$
$i \in I$ is a basis for a uniformity on $X$ ~\cite{Kelley}.  On
the other hand every uniformity can be constructed from a uniform
neighborhood system.
\\
The topology $ \tau_{U} $ defined by a uniformity $ U $, the
\textbf{uniform topology}, is constructed as follows:
$$ O(x) \in \tau_{U} \Longleftrightarrow \exists C \in U : St(x,C)
\subset O(x) $$ so $ \{ St(x,C) | x \in X , C \in U \} $ defines a
basis for the topology. The topology is Hausdorff if and only if $
\bigcap_{O(x) \in \tau_{U}} O(x) = \{x\} $ but it is not difficult
to see that this is equivalent with: $$\bigcap_{i \in I } B_{i}(x)
= \{x\}
$$ where $ \{ B_{i} (x) | i \in I, x \in X \} $ is the uniform
neighborhood system which generates $U$. To indicate the reasoning
followed in ~\cite{Bombelli} we will just state a few facts about
\textbf{quotient uniformities}.
\textbf{Terminology}
\begin{itemize}
\item Let $(X,U)$ and $(Y,V)$ be uniform spaces, a map $f : X
\rightarrow Y $ is \textbf{uniformly continuous} if and only if
$$ \forall C \in V : f^{ -1} ( C ) \in U $$
where for $ C = \{ A_{i} | i \in I \} $ , $ f^{-1} (C) = \{ f^{-1}
(A_{i}) | i \in I \} $.
\item A uniformity $ \tilde{U} $ on $X$ is finer than $U$ if and
only if every cover in $U$ belongs to $ \tilde{U} $.
\item Let $ \pi : X \rightarrow \tilde{X} $ be a surjective map
and $(X,U)$ a uniform space, the \textbf{quotient uniformity} $
\tilde{U} $ on $ \tilde{X} $ is the finest uniformity which makes
$ \pi $ uniformly continuous.
\end{itemize}
Notice that the existence of a quotient uniformity is guaranteed
by the lemma of Zorn, the uniqueness is immediate.  The obvious
question now is if $ \tau_{ \tilde{U}} $ is equal to the quotient
topology of $ \tau_{U} $.  The answer is in general no, but under
some special circumstances it works.
\begin{deffie}
A uniform neighborhood system $ \{ B_{i} (x) | x \in X, i \in I
\}$ is \textbf{compatible} with an equivalence relation on $X$ if
and only if
$$ \forall i \in I, x' \sim x \text{ and } y \in B_{i}(x)
\quad \exists y' \sim y : y' \in B_{i} ( x' )
$$
\end{deffie}
As envisaged, compatibility implies that $ \tau_{ \tilde{U}} $ is
equal to the quotient topology of $ \tau_{U} $.
\\
\begin{theo}
If $U$ is generated by $ \{ B_{i}(x) | i \in I, x \in X \} $ which
is compatible with $ \sim $  which is for example defined by a
surjective map, then the quotient uniformity $\tilde{U} $ on $
\tilde{X} = X / \sim $ is generated by the uniform neighborhood
system defined by:
$$ \tilde{B}_{i} ( \tilde{x} ) = \{ \tilde{y} | \exists x \in
\tilde{x} \text{ and } y \in \tilde{y} : y \in B_{i} ( x)\} $$ $
\forall \tilde{x} \in \tilde{X}, i \in I$.  Moreover $ \tau_{
\tilde{U}} $ is equal to the quotient topology of $ \tau_{U} $ and
a basis of neighborhoods of $ \tilde{x} \in \tilde{X} $ is $ \{
\tilde{B}_{i} ( \tilde{x} ) | i \in I \} $
\end{theo}
As mentioned, every uniformity can be generated by a family of
pseudodistances. In the case that the uniformity is generated by a
countable uniform neighborhood system, the topology is defined by
one pseudodistance, which is a distance when the uniformity is
Hausdorff.  Suppose $ C_{n} = \{ B_{n} (x) | x \in X \} $, $ n \in
\mathbb{N} $ , is a countable basis for a uniformity $U$, then we
can find a subsequence $ (n_{k})_{k} $ such that:
$$ \forall k, \quad w \in B_{n_{k}} ( z),  z \in B_{n_{k}} (y),  y \in B_{n_{k}} (x
) \Rightarrow  w \in B_{n_{k-1}} (x)$$ Assume $C_{n}$ is such a
basis.
\\
\begin{theo}
Let $ C_{n} $ be a countable basis of $U$, then with
$$ \rho (x,y) = \inf_{ \{ n \geq 0, y \in B_{n} (x) \} }
2^{-n} $$ the function
$$ d(x,y) = \inf_{ K \in \mathbb{N},   x_{k}} \sum_{ k = 1}^{K} \frac{1}{2} (
\rho (x_{k-1} , x_{k} ) + \rho ( x_{k} , x_{k-1} ) ) $$ is a
pseudodistance which generates $U$.  $ \{ x_{0} , \ldots , x_{K}
\} $ with $ x_{0} = x, x_{K} = y$ is a path in $X$. If $U$ is
Hausdorff then $d$ is a distance.
\end{theo}
Note that the function $d$ depends on the choice of basis $C_{n}$
and is therefore not canonical.  In the next chapter we apply this
machinery to the distance functions $d^{i}_{ \text{cau}},
d^{i}_{\text{vol}}$ and $ d^{i}_{ \text{geo} } $.
\section{A topology on $ \mathbb{L} (M) $ in the case $ M $ is compact}
In this section we first introduce the Schwartz topology on $
Diff(M) $, next we prove that $ f_{g, \tilde{g}}^{\text{cau}} $ is
continuous in this topology when $ g, \tilde{g} $ are globally
hyperbolic.  Then we give some characterizations of amenability
and give some hints why we think that large amenable subgroups $G$
can be found.  With a left invariant mean on $ L^{ \infty }(G) $
we can symmetrize the distance functions $d_{\text{cau}} ,
d_{\text{ vol}} , d_{ \text{geo} } $, these functions will define
a uniformity and thus a $G$ invariant pseudodistance on the space
of globally hyperbolic metrics.
\subsection{ A topology on $ Diff(M) $ }
We first state the result (in a slightly more general way as
needed) and then show the major lines of the construction
~\cite{Michor,Kriegl2}

\begin{theo}
Let $M$ and $N$ be ordinary manifolds with $M$ compact.  Then $
C^{ \infty} (M,N) $ has the structure of a $ C^{ \infty }_{c} $
manifold.  The local model at $ f \in C^{ \infty } ( M, N ) $ is
given by the nuclear Fr\'echet space $ C^{ \infty }_{f} ( M ,TN )
$
\end{theo}
One proves that $ Diff(M) $ is a $ C^{ \infty }_{c} $ open
submanifold of $ C^{ \infty } (M,M) $, moreover $ Diff( M ) $ is a
$ C^{ \infty }_{c} $ Lie group.  It is also proven that the
composition and inversion are continuous.
\\ First we recall when a map $f$ is $ C^{ \infty }_{c}$.
Let $E, F$ be locally convex Hausdorff linear spaces and $W
\subset E$ be open, then $f : W \rightarrow F $ is $ C^{1}_{c} $
if and only if there exists a linear mapping $ Df : E \rightarrow
L(E,F) $ such that $$ \lim_{ t \rightarrow 0 } \frac{ f(x + tv) -
f(x) }{t} = Df(x)v \quad \forall v \in E, \quad x \in W, \quad t
\in \mathbb{R}  $$ such that the mapping $ W \times E \rightarrow
F : (x,y) \rightarrow Df(x)y$ is continuous.  The set of all
$C^{1}_{c} $ mappings is a linear space.  The space $ C^{k}_{c} $
is defined by recursivity: a map $f$ is $C^{k}_{c} $ if $ D^{k-1}f
: W \times E^{k-1} \rightarrow F $ is $ C^{1}_{c} $. Finally $$
C^{ \infty }_{c} = \bigcap_{k \geq 1 } C^{ k }_{c} $$ Now for
Fr\'echet spaces this concept of smoothness is equivalent to the
following: a map $ f : W \subset E \rightarrow F $ is smooth if
and only if every $ C^{ \infty }_{c} $ curve on $W$ is mapped to a
$ C^{ \infty }_{c} $ curve on $F$. Many results , which we will
state now, can be generalized to a larger class of locally convex
Hausdorff linear spaces ~\cite{Balanzat}.  Let $  W \subset
\mathbb{R}^{n} $, then the topology on $ C^{ \infty } ( W, F  ) $
where $F$ is an euclidian space, is defined by the family $
\rho_{m,K }$ of seminorms. Let $m \in \mathbb{N}_{0} $, $ p =(
p_{1},p_{2}, \ldots , p_{n} ) \in \mathbb{N}^{n}$ and $ |p| =
\sum_{i=1}^{n} p_{i} $, then for any compact $K \subset W$
$$ \rho_{m,K} (f) = sup_{ |p| \leq m } (  sup_{ x \in K } \parallel
D^{p}f(x) \parallel ) $$ where $$ D^{p} = \frac{ \partial^{ |p|
}}{ \partial x_{1}^{ p_{1}} \partial x_{2}^{ p_{2}} \ldots
\partial x_{n}^{ p_{n}} } $$  It is well known that this family of
seminorms makes $ C^{ \infty } ( W , F ) $ into a nuclear
Fr\'echet space  ~\cite{Dubinsky,Yosida}.  Now let $M$ be any
ordinary manifold and $( U ,\phi )$ be a chart, the map
$$ \phi^{*} : C^{ \infty } ( U, F) \rightarrow C^{ \infty }( \phi
( U ) , F ) : f \rightarrow f  \circ \phi^{-1} $$ is a linear
isomorphism and induces on $ C^{ \infty } ( U, F) $ the structure
of a nuclear Fr\'echet space.  Because $M$ is second countable
there exists a countable covering of charts $ ( U_{ k } , \phi_{ k
} ) $ such that one can construct the following restriction maps
$$ C^{ \infty } ( M, F ) \rightarrow C^{ \infty } ( U_{k}, F ) $$
which implies that
$$ C^{ \infty } ( M, F ) = \lim_{k} C^{ \infty } ( U_{k}, F  )$$
where this limit is a projective limit of nuclear Fr\'echet spaces
and is hence a nuclear Fr\'echet space
~\cite{Abbati,Yosida,Dubinsky}. This topology is known as the
Schwartz topology.  One can introduce jet bundles in order to
characterize the Schwartz topology, the reader is referred to
~\cite{Kriegl2}.
\\ Let $ exp : U \subset TN \rightarrow N $ be the
exponential map associated to a Riemannian metric $h$ on $N$ , $
U$ an open neighborhood of the zero section on which the
exponential map is defined and $ \pi_{N} : TN \rightarrow N $ the
canonical projection.  One can choose $U$ such that
$$ \nu \equiv ( \pi_{N} , exp ): U \rightarrow N \times N : v \rightarrow ( \pi_{N} (v) ,
exp_{ \pi_{N} (v)}(v) ) $$ is a diffeomorphism from $U$ to a
neighborhood $V$ of the diagonal in $N \times N$.  Now we can
define the local model $C^{ \infty }_{f} ( M, TN ) $ of $ f \in
C^{ \infty } ( M, N ) $:
$$ C^{ \infty }_{f} ( M, TN ) = \{ g \in C^{ \infty } (
M, TN ) | \quad \pi_{N} \circ g = f \} $$ $C^{ \infty }_{f} ( M,
TN ) $ is a linear space which can be identified with $ Sec( E ) $
in the bundle $( E , \pi_{E} ,M )$ where $E$ is defined as
follows:
$$ E = \{ (x,v) | v \in TN_{f(x)} \} $$
The topology on $E$ is defined by the open sets $ O_{W,U} = \{
(x,v_{f(x)}) | \quad x \in W \quad \text{and} \quad v_{f(x)} \in U
\} $ where $W$ is open in $M$ and $U$ is open in $TN$ with $f(W)
\subset \pi_{N} ( U ) $. The differential structure is defined by
the charts $ ( O_{W,U} , ( \chi , pr_{2} \circ \psi  )) $ where $
( W ,\chi ) $, $ ( U , \psi  ) $ are charts in $ M $ respectively
$ TN $, with $ \psi( U ) \subset V \times F $ and $ pr_{2} $ is
the projection on the second factor.  We clearly can endow the
$C^{ \infty } $ sections on $E$ with the Schwartz topology, so
this induces on $C^{ \infty }_{f} ( M, TN ) $ the structure of a
nuclear Fr\'echet space.
\\ We can now define the charts on $ C^{ \infty } ( M, N )
$.  Define $$ U_{f} = \{ g \in C^{ \infty } ( M, N ) | \quad
(f(x),g(x)) \in V \quad \forall x \in M \} $$ This induces the
map:
$$ u_{f} : U_{f} \rightarrow C^{ \infty }_{f} ( M , TN ) $$ with $ u_{f} (g) (x) =
exp_{f(x)}^{-1} ( g(x)) =  ( \nu^{-1} \circ ( f \times g )) ( x )
$. \\ $u_{f}$ maps $U_{f}$ bijectively to $ \{ s \in C^{ \infty
}_{f} ( M , TN ) | s(x) \in U \} $ which is by definition open in
the Schwartz topology because $M$ is compact.  Now the inverse of
$u_{f}$ is given by:
$$ u_{f}^{ -1 } ( s )(x)  =  exp_{f(x)}( s(x) ) $$ so $ u_{f}^{-1}
(s) = ( pr_{2} \circ \nu ) \circ s $.  The atlas on $ C^{ \infty }
( M , N ) $ is hence defined as $$ \{ ( U_{f} , u_{f} ) | f \in
C^{ \infty } ( M, N ) \} $$  One calculates the transition maps
and proves that they are $ C^{ \infty }_{c} $.  It is clear that
if $ U_{f} \cap U_{ g } \neq \emptyset $ that $ u_{f} \circ
u_{g}^{ -1 } : u_{g} ( U_{f} \cap U_{ g } ) \rightarrow u_{f} (
U_{f} \cap U_{ g } ) $ is given by :
$$ (( u_{f} \circ u_{g}^{ -1 } )( s ))(x)  = u_{f(x)}( exp_{g(x)}
( s(x) )) = exp_{f(x)}^{-1}( exp_{g(x)} ( s(x) )) $$ or $$ u_{f}
\circ u_{g}^{ -1 } = exp_{f}^{-1} \circ exp_{g} $$  This
transition map is $ C^{ \infty }_{c} $ if and only if they map
smooth curves to smooth curves.  But the smooth curves of $ C^{
\infty }_{f} ( M , TN ) $ correspond with the smooth sections of
the bundle $$ \mathbb{R} \times E \rightarrow \mathbb{R} \times M
$$ which are preserved by transition maps.  It is easy to see that
the differential structure is independent of the chosen Riemannian
metric so we arrive at our result.  It is worthwhile noticing that
$ exp $ is defined everywhere on $TM$ and that in $M$ every two
points can be connected by a unique $h$ - geodesic (this is a
special case of the Hopf - Rinow theorem) ~\cite{Jost}.
\\ \\
\begin{theo}
Let $g, \tilde{g} $ be globally hyperbolic, the map $ f_{g,
\tilde{g} }^{\text{cau}} $ is continuous in the Schwartz topology
on $Diff(M)$
\\
Proof
\end{theo}
In this proof $\bar{d}$ denotes the distance on $M$ corresponding
with the Riemannian metric $h$; we shall also use the shorthand
notation $V^{2} \equiv V \times V$.  We will prove the upper
continuity, the lower continuity is identical.  It is clearly
sufficient to prove the continuity of $ f_{g \tilde{g}
}^{\text{cau}} $ in the identity diffeomorphism, for every other
diffeomorphism the proof repeats almost ad verbatim.  The
Alexandrov sets of $ \phi^{*} \tilde{g} $ satisfy the following
property : $$ A_{ \phi^{*} \tilde{g} } (p,q) = \phi^{-1} (
\tilde{A} ( \phi (p) , \phi ( q ) ) ) $$ This implies that:
\begin{equation}  \alpha_{g \phi^{*} \tilde {g} } (p, q)  =  \frac{ V ( A
(p,q )
 \bigtriangleup \phi^{-1} (  \tilde{A} ( \phi
(p) , \phi ( q ) ))) }{ V(A(p,q) \cup \phi^{-1} (  \tilde{A} (
\phi (p) , \phi ( q ) ))) } \end{equation} We introduce the
following notations:
\begin{itemize}
\item $\mathcal{N} = \{ (p,q) | q \in J^{+}(p) \setminus \tilde{J}^{+}(p)
\}$
\item $\tilde{\mathcal{J}}^{+} = \{ (p,q) | q \in \tilde{J}^{+}(p)
\}$
\item $ \mathcal{O}_{\phi} = \{(p,q) | q \notin \tilde{J}^{+} (p)
\text{ but } \phi(q) \in \tilde{J}^{+} ( \phi(p)) \} $
\end{itemize}
It is clear that $\alpha_{g \phi^{*} \tilde{g}} (p,q) > 0$ implies
that $(p,q) \in \mathcal{N} \cup \mathcal{O}_{\phi} \cup
\tilde{\mathcal{J}}^{+}$.  This implies that \begin{equation}
d_{\text{cau}}(g, \phi^{*} \tilde{g}) \leq V^{2}(\mathcal{N}) +
V^{2}( \mathcal{O}_{\phi} ) + \int_{\tilde{\mathcal{J}}^{+}}
\alpha_{g \phi^{*} \tilde{g}}(p,q) dV(p) dV(q) \end{equation}
Choose $\epsilon>0$, we show first that the second term on the
right hand side can be made smaller than $\frac{\epsilon}{4}$. For
this purpose we define for every $X \subset M$ the set $X_{\zeta}
= \{ y \in M \setminus X | \quad \exists x \in X : \bar{d}(x,y) <
\zeta \}$. The reader can check that the mapping $(\zeta, p)
\rightarrow V( (\tilde{J}^{+}(p))_{\zeta})$ is continuous.  Hence
there exists a $\delta_{0}$ such that $$V^{2}( \bigcup_{p \in M}
\{p\} \times ( \tilde{J}^{+}(p))_{\delta}) < \frac{\epsilon}{4}$$
for all $\delta < \delta_{0}$.  If $\phi \in C^{\infty}_{id} (M,
TM)$ satisfies the property that $\forall p \in M:
\bar{d}(p,\phi(p))< \delta_{0}$ then one has that:
$$ \mathcal{O}_{\phi} \subset \bigcup_{p \in M} \{p\} \times (
\tilde{J}^{+}(p))_{\delta_{0}}$$ and hence
$$ V^{2}(\mathcal{O}_{\phi}) < \frac{\epsilon}{4} $$
The condition on $\phi$ determines an open subset $U_{\delta_{0}}
\subset C^{\infty}_{id}(M,TM)$.  In order to make further
estimations we have to split $\tilde{\mathcal{J}}^{+}$ in two
parts.  \\
Let $ \beta_{p} : \tilde{J}^{ + }(p) \rightarrow
\mathbb{R} : q \rightarrow V( \tilde{A} ( p, q)) $ and $ S^{
\delta }_{p} = \{p\} \times \{ q \in \tilde{J}^{+}(p) | \beta_{p}
( q ) \leq \delta \} $.  It is proven in appendix B that there
exists a $ \delta_{1} > 0 $ such that $ V^{2}( \cup_{p \in M} S^{
\delta_{1} }_{p} ) \leq \frac{\epsilon}{2} $.  If we use the
notation that $M_{ \delta_{1} } = \{ ( p,q ) | V(\tilde{A}(p,q))
\geq \delta_{1}  \} \subset M \times M $ then (2) becomes:
\begin{equation} d_{\text{cau}} ( g , \phi^{*} \tilde{g} ) \leq
\frac{3 \epsilon}{4} + V^{2}( \mathcal{N} ) +
\int_{M_{\delta_{1}}} \alpha_{g \phi^{*} \tilde{g} } (p,q) dV(p)
dV(q) \end{equation} If we show that there exists an open subset
$U_{\delta_{1}} \subset C^{\infty}_{id}(M,TM)$ such that for all $
X \in U_ { \delta_{1} }$ and $ \phi = exp_{id} ( X ) $:
\begin{equation} \int_{M_{\delta_{1}}} \alpha_{g \phi^{*}
\tilde{g} } (p,q) dV(p) dV(q)  < \int_{M_{\delta_{1}}} \alpha_{g
\tilde{g} } (p,q) dV(p) dV(q) + \frac{\epsilon}{4} \end{equation}
then
$$ d_{\text{cau}} ( g , \phi^{*} \tilde{g} ) < \epsilon +
V^{2}(\mathcal{N}) + \int_{M_{\delta_{1}}} \alpha_{g \tilde{g} }
(p,q) dV(p) dV(q) < \epsilon + d_{\text{cau}}(g, \tilde{g}) $$
 On $M_{\delta_{1}}$ one can easily bound $\alpha_{g \phi^{*}
\tilde{g} }(p,q)$ by:
\begin{equation} \frac{ V( A(p,q) \bigtriangleup \tilde{A}(p,q) )
+ V( \tilde{A}(\phi(p) ,\phi(q)) \bigtriangleup \tilde{A}(p,q) ) +
V( \phi^{-1} ( \tilde{A} ( \phi (p) , \phi ( q) ) \bigtriangleup
\tilde{A}( \phi( p) , \phi (q) ) )}{ V( A(p,q) \cup \tilde{A} (
p,q) ) - V( \phi^{-1} ( \tilde{A} ( \phi (p) , \phi (q ) ) )
\bigtriangleup \tilde{A} (p,q) )}
\end{equation}
In order to make the final estimates, we need two lemma's.
\newtheorem{lem}{lemma}
\begin{lem}
$\forall \kappa > 0 \quad \exists U_{ \kappa } \subset C^{ \infty
}_{id} ( M , TM ) $ such that $ X \in U_ { \kappa }$ implies that
for $ \phi = exp_{id} ( X ) $:
$$ V( \tilde{A}( p,q) \bigtriangleup \tilde{A}( \phi (p) ,\phi
(q)) ) < \kappa V ( \tilde{A}( p,q) ) $$ $ \forall (p,q) \in M_{
\delta_{1}} $. \\
\\
Proof
\end{lem}
Define the function $$ F : M^{2}_{ \delta_{1}} \rightarrow
\mathbb{R} : ((p,q),(r,s)) \rightarrow \frac{ V( \tilde{A}( p,q)
\bigtriangleup \tilde{A}( r ,s ))}{ V( \tilde{A} (p,q)) } $$ In
appendix C it is proven that $F$ is continuous.  This proves that
$ \exists \delta < \delta_{0}$ such that $\forall (p,q) \in M_{
\delta_{1}}$ and $p',q'$ such that $\bar{d}(p,p'), \bar{d}(q,q') <
\delta$:
$$ V( \tilde{A}( p,q) \bigtriangleup \tilde{A}(p' ,q') ) <
\kappa V ( \tilde{A}( p,q) ) $$ $ \bigcup_{p \in M } \{p\} \times
B(p, \delta) $ clearly determines an open neighborhood of the
diagonal in $ M^{2}$ and this yields the open neighborhood $ U_{
\kappa } \subset U_{\delta_{0}} $.  $\square$
\\
\begin{lem}
$ \forall \kappa  > 0 $, then there exists a neighborhood
$W_{\kappa}$ of the zero section such that $ \forall (p,q) \in M_{
\delta_{1}} $ we have that  :
$$ V(\tilde{A} (\phi(p),\phi(q)) \bigtriangleup \phi^{-1}
\tilde{A} ( \phi (p) , \phi (q) ) ) < \kappa V( \tilde{A} ( p ,q
))
$$ $ \phi \in W_{\kappa}$ implies that $ d_{vol} ( \phi_{*} g , g ) < ln(3) $.
\end{lem}
\begin{rem}
The condition that $ d_{vol} ( \phi_{*} g , g ) < ln(3) $ makes
restrictions on $ \bar{d}(r, \phi ( r )) \quad \forall r \in M $
and on the determinant of the Jacobian of $ \phi $ in any chart of
the cover $ U_{k} $.  Choose $p \in M$ and charts $ ( U_{p} , \chi
) , ( U_{ \phi ( p )}, \nu ) $ belonging to the cover, we denote
the coordinates in $U_{p} $ with $ x^{ \alpha }$ and the
coordinates in $ U_{ \phi (p) } $ with $ y^{ \beta } $.  We note $
\phi^{ \beta } ( x^{ \alpha } ) = y^{ \beta } ( \nu \circ \phi
\circ \chi^{-1} ( x^{ \alpha } ) ) $, so we can estimate :
$$ \mid ln \left(  \frac{ \sqrt{ - \mid \phi_{*}g( \phi^{\beta } (x^{ \alpha } )) \mid } }
{ \sqrt{ - \mid g( \phi^{ \beta} (x^{ \alpha } ) ) \mid } }
\right) \mid $$ by
$$ \mid ln \left( \mid \frac{ \partial x^{ \alpha }}{ \partial \phi^{ \beta } ( x^{ \nu } )} \mid
\right) \mid + \mid ln \left( \frac{ \sqrt{ - \mid g(x^{ \alpha })
\mid } }{ \sqrt{ - \mid g( \phi^{ \beta } ( x^{ \alpha } )) \mid }
} \right) \mid $$ \\
These restrictions determine open sets
in $ C^{ \infty }_{id } ( M , TM )$ so the condition is justified.
\end{rem}
The proof of lemma 2 is given in appendix D.  The result of
proposition $8$ follows now immediately.  Lemma $1$ and $2$
determine an open neighborhood $W_{\kappa} \cap U_{\kappa}$ of the
identity such that
$$ V(\tilde{A} (\phi(p),\phi(q)) \bigtriangleup \phi^{-1} \tilde{A}
( \phi (p) , \phi (q) ) ) < \kappa V( \tilde{A} ( p ,q ))$$ and
$$  V(\tilde{A}
(\phi(p),\phi(q)) \bigtriangleup \tilde{A} ( p , q) ) < \kappa V(
\tilde{A}( p,q))$$ Substitution of these inequalities in (5)
implies that $ \alpha_{g ( \phi^{*} \tilde{g})} (p,q) $ can be
bounded on $ M_{ \delta_{1}} $ by:
$$ \frac{1}{1 - 2 \kappa } \alpha_{g \tilde{g}} (p,q) + \frac{2
\kappa }{ 1 - 2 \kappa } $$ This can be bounded again by
$$ \alpha_{g \tilde{g}} (p,q) + \frac{ 4
\kappa }{ 1 - 2 \kappa } $$ Putting this in inequality (4) we
obtain that $ \kappa < \frac{  \epsilon }{ 16 \tilde{M}^2 + 2
\epsilon } $ yields the result.  $ \square $
\\
\\
The proof is lengthy because of the construction of $M_{
\delta_{1}} $, which is the key idea of the proof.  One has to
avoid to come arbitrarily close to the boundary of the lightcone
because long skinny Alexandrov sets will force the diffeomorphism
to become very small.  This situation can be restored in the
compact case, this is obviously impossible when $M$ is not
compact! A much more interesting - and difficult - question is
under which causal restrictions on $g, \tilde{g}$ the mapping
$f_{g \tilde{g} }^{ \text{cau}}$ is measurable in the Schwartz
topology. We didn't find an answer to this question yet.  It is
obvious that the function $f_{g , \tilde{g}}^{ \text{vol}} $ is
continuous in the Schwartz topology. The continuity of $f^{
\text{geo}}_{g , \tilde{g}} $ needs a bit of explanation.  Choose
$ \epsilon > 0$ , we have for all $ \psi \in Diff(M) $ that: $$ |
d_{ \text{geo}} ( g, \tilde{g} ) - d_{ \text{geo}} ( g, \psi^{*}
\tilde{g} )| \leq
 d_{ \text{geo}} ( \tilde{g} , \psi^{*} \tilde{g} )$$
Using that $ \lambda_{ \psi^{*} \tilde{g} } ( p, q ) = \tilde{
\lambda } ( \psi (p) , \psi(q) ) $  we only have to prove that
there exists an open neighborhood $V$ of the identity
diffeomorphism in $ Diff(M) $ such that $ \psi \in V$ implies that
$$ | \tilde{\lambda}( \psi (p ) , \psi ( q ) )
- \tilde{ \lambda} ( p, q ) | < \epsilon $$  One can easily proof
this using the continuity of $(p,q) \rightarrow \tilde{ \lambda} (
p, q )$ and applying a doubling trick such as in the proof of
lemma $2$.

\subsection{Amenability }

The aim of this paragraph is to introduce the reader to the
concept of amenability.  It contains no new results.  The
intention is to give the reader an idea of the technical
difficulties which show up.  We will first state some of the main
results for general topological groups; later on we give a list of
equivalent characterizations in case of locally compact
topological groups or Lie groups.  For more concise treatments,
the reader is referred to ~\cite{Greenleaf,Higgins,Keller}.
\\ Let G be a topological group and $X $ a closed subspace
of $ L^{ \infty } (G ) $.  $X$ is assumed to be one of the
following subspaces:
\begin{itemize}
\item $ X = L^{ \infty } ( G )$ \\
\item $ X = CB ( G ) $ ( continuous bounded functions )
\end{itemize}
A mean $m$ on $X$ is a positive element of the dual $X^{*}$ such
that
$$m(1) = 1 = \parallel m \parallel $$
This condition is equivalent with the requirement that
$$ \inf_{x \in G} f(x) \leq m(f) \leq \sup_{ x \in G} f(x) \quad
\forall f \in X$$  Now we are especially interested in invariant
means (IM).  To define invariance we first define the right and
left translations $ f_{g} $ respectively $f^{g}$ of a function $ f
\in X$ by an element $ g \in G $ : $$ f_{g} ( x ) = f( x g ) $$
and $$ f^{g} (x) = f(g^{-1} x )$$ It is clear that $$ f_{ gh } ( x
) = f( x g h ) = ( f_{h} )_{g} ( x )
$$ so right translation defines a representation of $G$ on $X$ and the same is true for the left
translation.  A RIM $m$ is a mean which satisfies :
$$ m(f_{x} ) = m(f) \quad \forall x \in G, f \in X $$
and left invariant means (LIM) are defined in the same way.  An IM
is a LIM and RIM.  It is obvious that an IM on $ X = L^{ \infty }
( G )$ is also a IM on $ X = CB ( G ) $, the converse however is
not true in general. However when $G$ is locally compact the two
notions are the same  ~\cite{Greenleaf,Pier}. Because $Diff(M)$ is
not locally compact and considering proposition $8$ it is obvious
that our purpose is to consider an IM on $ X = CB ( G ) $. Because
the inversion and multiplication are continuous in the Schwartz
topology on $ Diff(M )$, the existence of a LIM (RIM) on $ CB (G
)$ where $G$ a l.c. subgroup of $ Diff(M) $ implies the existence
of an IM (~\cite{Greenleaf} page $2$, ~\cite{Pier} p. $36$). We
will call groups which have a IM on the space of continuous
bounded functions amenable (this is just our convention, the word
amenable is also used in other meanings).
\\ $ CB ( G ) $ endowed whith the sup norm is a Banach
space.  It is well known that the unit ball in $X^{*}$ is weak$*$
compact, hence it is obvious that the convex set of IM is weak$*$
compact.  This implies that if $G = \cup_{ \alpha \in I } H_{
\alpha } , \quad I$ a directed set and $ H_{ \alpha } $ a net of
closed amenable subgroups, we have that $G$ is amenable
(~\cite{Greenleaf} p. $30$). The lemma of Zorn implies then that
we can find maximal amenable subgroups.  Now the most interesting
properties and characterizations arise when $G$ is locally
compact. Intuitively most locally compact groups are amenable, it
is a disadvantage however that a RIM is not unique.  It is also
important to point out that the right invariant Haar measure is
not a RIM unless $G$ is compact, this is logical since a mean of a
function of compact support is zero when $G$ is not compact. For
more information concerning amenable l.c. groups we refer the
reader to appendix E.  There it is shown that a great deal of l.c.
groups are amenable, so one could raise the question if in
$Diff(M)$ one can find "large" maximal amenable subgroups. "Large"
means for example that the subgroup has infinite dimension.  This
kind of questions are not dealt with yet.
\\
In the following paragraphs we always assume that $G$ is a maximal
closed amenable subgroup of the diffeomorphism group.

\subsection{ A $G$ - invariant uniformity and pseudodistance on
the space of globally hyperbolic metrics} Now we return to the
results starting at section $3$.  We call two globally hyperbolic
metrics $g , \tilde{g} $ on a compact space time $ ( \zeta ,
\epsilon , \alpha ) $ close if and only if :
$$ \begin{cases}
d_{ \text{ vol }}(g , \tilde{g}) \leq \zeta \\
d_{ \text{cau }}  ( g, \tilde{g} ) \leq \epsilon \\
d_{ \text{geo} } (g , \tilde{g} ) \leq \alpha
\end{cases} $$
We say that $ ( \zeta' , \epsilon' , \alpha' ) < ( \zeta ,
\epsilon , \alpha  ) $ if and only if $ \zeta' < \zeta , \epsilon'
< \epsilon , \alpha' <  \alpha $. This turns $ \mathbb{R}^{3} $
into an oriented net.  Define now:
$$ B_{ \zeta ,\epsilon , \alpha } (g) = \{ \tilde{g} | (
g , \tilde{g} ) \quad \text{are} \quad ( \zeta' , \epsilon' ,
\alpha' ) \quad \text{close with} \quad ( \zeta' , \epsilon' ,
\alpha' ) < ( \zeta , \epsilon , \alpha ) \}
$$ Proposition $1$ and $2$  and the following remarks prove that
the $B_{ \zeta ,\epsilon , \alpha  } (g)$ form a uniform
neighborhood system.  So we know that the covers $$ C_{ \zeta
,\epsilon , \alpha  } = \{ B_{ \zeta ,\epsilon , \alpha  } (g) | g
\in \mathbb{L} ( M ) \} $$ form a basis of a uniform neighborhood
system $U$.  Clearly we have a countable subbasis so according to
proposition $6$  there exists a diagonally invariant pseudometric
$d$.  However there is no guarantee that $d$ is bounded, therefore
we introduce a cut off scale $ \eta $.  A modified version of
proposition $6$ gives the following result.
\\
\begin{theo}
Let $ C_{n} = \{ B_{n} ( g ) | g \in \mathbb{L} ( M ) \}  $ be a
countable basis of $U$ which satisfies the property that :
$$ \forall n, \quad g \in B_{n} ( g_{1}),  g_{1} \in B_{n} (g_{
2}),  g_{2} \in B_{n} (\tilde{g}) \Rightarrow  g \in B_{n-1}
(\tilde{g})$$ and let $ \eta > 0 $ be a cut off scale. Put
$$ \rho (g, \tilde{g} ) = \inf_{ \{ n(g, \tilde{g}) \geq 0, \tilde{g} \in B_{n} (g) \} }
2^{-n(g, \tilde{g})}  $$ then we have that the function
$$ d(g,\tilde{g}) = \min \{ \inf_{ K \in \mathbb{N},   x_{k}} \sum_{ k = 1}^{K} \frac{1}{2} (
\rho (x_{k-1} , x_{k} ) + \rho ( x_{k} , x_{k-1} ) ) , \eta \}
$$ is a diagonally diffeomorphism invariant pseudodistance which
generates $U$ - $ \{ x_{0} , \ldots , x_{K} \} $ with $ x_{0} = g,
x_{K} = \tilde{g}$ is a path in $ \mathbb{L} ( M )$. Moreover $d$
is a distance on the space of strongly causal metrics.
\\
Proof
\end{theo}
The fact that $d$ is a distance on the space of strongly causal
metrics follows from proposition $3$.  $\square $
\\
One could have esthetical objections against this cut off scale
from the mathematical point of view.  For a physicist however,
this cut off scale is quite natural, one is a priori only
interested in these space times $ \tilde{g} $ which are a "good
approximation" of $g$.  Moreover the cut off doesn't change the
topology generated by $d$.
\\ Let $g, \tilde{g} $ be globally hyperbolic, we show now
that $ f_{g , \tilde{g}}$ is continuous in the Schwartz topology.
Choose $ \epsilon > 0$, from the triangle inequality we get that:
$$ | d( g , \psi^{*} \tilde{g} ) - d(g, \tilde{g} ) | \leq d(
\tilde{g} , \psi^{*} \tilde{g} ) $$ Proposition $9$ implies that
there exists a $n_{0}$ such that $ \forall n \geq n_{0} : B_{n} (
\tilde{g} ) \subset B_{ \epsilon }^{d} (\tilde{g} ) $ where $ 2^{-
n_{0}} < \epsilon $.  The continuity of $f_{ \tilde{g}  \tilde{g}
}^{ \text{cau}}, f_{g, \tilde{g}}^{\text{vol}} $ and $f_{g,
\tilde{g}}^{\text{geo}}$ implies that there exists an open
neighborhood $V$ of the identity in $Diff(M)$ such $ \psi \in V$
implies that $ \psi^{*} \tilde{g} \in B_{n_{0}} ( \tilde{g} ) $.
This proves the continuity.
\\
As stated in the introduction :
$$ \tilde{d} : ( g, \tilde{g} ) \rightarrow \begin{cases} A(f_{g, \tilde{g}} + f_{\tilde{g} , g}) \quad \text{ when
} \quad \tilde{g} \neq \psi^{*} g
\quad \forall \psi \in Diff(M) \\
0 \quad \text{ otherwise } \end{cases}
$$  (with $A$ the $G$ invariant mean) is a $G$ invariant pseudodistance on the space of globally
hyperbolic metrics.  Now we make some remarks concerning the
Hausdorff character of $ \tilde{d} $.  Obviously one has:
$$ \tilde{d} ( g , \tilde{g} ) = 0 \Rightarrow \inf_{ \phi \in
Diff(M) } d ( g , \phi^{*} \tilde{g} ) = 0$$ but the implication
doesn't necessary go the other way.  This is clearly an advantage
in comparison with the construction given in
~\cite{Bombelli,Bombelli2}. On the other hand, in the same paper
one obtained a fully diffeomorphism invariant pseudodistance,
whereas ours is only fully $G$ invariant. There remains the
question what size this $G$ has, when for example under some
topological restrictions on $M$ the Lie algebra of $G$ has
infinite cardinality.  In example 3 one has a group $G \subset
Diff(M)$ such that $\inf_{ \phi \in G } d ( g , \phi^{*} \tilde{g}
) = 0$ but $ \tilde{d} ( g , \tilde{g} )
> 0$.
\\
Let $g, \tilde{g} $ be globally hyperbolic, as anticipated before
on page 7, we investigate the continuity of $ \epsilon \rightarrow
d_{ \epsilon} ( g , \tilde{g} ) $ (we have dropped the index "i"
in the notation).  This function is clearly left continuous.  This
can be seen as follows; choose $\epsilon > 0$ and $ \kappa > 0$,
and suppose there exists a sequence $p_{n} , q_{n} $ such that $
V(A(p_{n} , q_{n} )) \geq \epsilon - \frac{1}{n}, \quad
V(\tilde{A}(p_{n} , q_{n} )) \geq \epsilon - \frac{1}{n} $ and
$$ \frac{\lambda(p_{n},q_{n}) }{\tilde{\lambda}(p_{n},q_{n})} \leq
e^{ - ( d_{ \epsilon} ( g, \tilde{g} ) + \kappa ) } $$ or
$$ \frac{\lambda(p_{n},q_{n}) }{\tilde{\lambda}(p_{n},q_{n})} \geq
e^{  d_{ \epsilon} ( g, \tilde{g} ) + \kappa  } $$ Without loss of
generality we can assume the latter and because $M$ is compact we
can find accumulation points $p$ and $q$ such that:
$$ \frac{\lambda(p,q) }{\tilde{\lambda}(p,q)} \geq
e^{  d_{ \epsilon} ( g, \tilde{g} ) + \kappa  } $$ and $ V(A(p ,
q) ) \geq \epsilon , \quad  V(\tilde{A}(p , q)) \geq \epsilon  $
which is a contradiction. This proves our claim.  In general one
does not have right continuity and it is not so difficult to find
counterexamples. One even can construct counterexamples where $M$
is compact and $ \epsilon \rightarrow d_{ \epsilon} ( g ,
\tilde{g} ) $ is not continuous in a countably infinite number of
points.  This is due to the non locality of the definition of $
d_{ \epsilon } $.

\section{The noncompact case}

We will now make a similar construction as before, however some
more care is needed.  There are 3 obvious difficulties:
\begin{itemize} \item It is clear that the functions
$d^{i}_{\text{vol}} , d^{i}_{ \text{cau }} $ and $ d^{i}_{
\text{geo}} $ are \underline{not} diagonally diffeomorphism
invariant when one considers diffeomorphisms on $M$.  They are
however invariant with respect to the subgroup of all
diffeomorphisms $ \phi $ such that $ \phi_{ | W_{i} } \in
Diff(W_{i})  $.
\item If $g, \tilde{g} $ are globally hyperbolic on $M$, one could
be tempted to consider the restrictions of these metrics to
$W_{i}$. It is however meaningless to look at $W_{i}$ as a space -
time by itself because the restrictions are not necessarily
globally hyperbolic. So the causal relations we consider on $W_{i}
\times W_{i} $ are of the type $ \prec_{|W_{i}} $ and not $
\prec_{ W_{i} } $.  In that way we preserve the qualities of
global hyperbolicity we needed to prove theorem $8$.
\item When $M$ is not compact, the Schwartz topology on $Diff(M)$
gets more complicated.  One has that $Diff(M)$ is an open
$C^\infty _c$ submanifold of $C^\infty _{\mathcal{F} \mathcal{D}}
(M,M)$, composition and inversion are smooth.  The Lie algebra of
the smooth infinite dimensional Lie group $Diff(M)$ is the
convenient vector space $\frak{X} _{cpt}(M)$ of all smooth vector
fields on M with compact support, equipped with the negative of
the usual Lie bracket and with the $\mathcal{D}$ topology.  The
exponential mapping $\exp : \frak{X}_{cpt}(M) \to Diff(M)$ is the
flow mapping to time 1, and it is smooth.  A definition of the $
\mathcal{F} \mathcal{D} $ topology can be found in
~\cite{Kriegl2}.
\end{itemize}
As in the previous chapter we make of $ \mathbb{N}_{0} \times
\mathbb{R}^{3} $ a directed net as follows:
$$ ( i', \zeta' , \epsilon' , \alpha' ) < ( i , \zeta ,
\epsilon , \alpha  )
$$ if and only if $$ i' >  i, \zeta' < \zeta , \epsilon' <
\epsilon , \alpha' < \alpha $$  In the same way one can define a
countable basis of neighborhoods $B_{n} (g) = B_{ n, \frac{1}{n} ,
\frac{1}{n} , \frac{1}{n} } ( g ) $ as follows:
$$ B_{n} (g) = \{ \tilde{ g} | ( g , \tilde{g} ) \quad  \text{are}
\quad (  i', \zeta' , \epsilon' , \alpha' ) \quad \text{close
with} \quad (  i', \zeta' ,  \epsilon' , \alpha' ) < ( n,
\frac{1}{n} ,  \frac{1}{n} , \frac{1}{n}  ) \} $$ It is clear that
the uniform topology constructed here doesn't depend on the choice
of the sequence $W_{i}$. Suppose one has two sequences $W_{i}$ and
$V_{j}$, then for every $i$ there exists a $j$ such that $W_{i}
\subset V_{j} $.  It is obvious then that for all $ \zeta ,
\epsilon , \alpha $ one has that $B^{V}_{  j , \zeta , \epsilon ,
\alpha } ( g ) \subset B_{ i , \zeta , \epsilon , \alpha } ( g )
\quad \forall g \in \mathbb{L}(M) $ and one can make the
same reasoning for the $ V_{j} $.   \\
We show now that the topology defined by the $B_{n}(g) $ does not
depend on the function $f$ we defined in the beginning.  Our
restrictions on $f$ made sure that 3 out of 4 defining properties
for a uniform neighborhood system are satisfied. However the
second defining property can be broken. This can be seen as
follows; if $g$ and $\tilde{g}$ are $( i , \zeta , \epsilon ,
\alpha )$ close then they are $( j , \zeta , \frac{f(V(W_{i})}{
f(V(W_{j}))} \epsilon , \alpha )$ close $ \forall j \leq i$.  As
one notices there is a volume dependent factor $\frac{f(V(W_{i})}{
f(V(W_{j}))}$ in the formula, which makes the partial order $<$
dependent on the volume scale of $g$, and hence $<$ cannot define
a uniform neighborhood system.  The most obvious thing to do is to
redefine $<$ as $ \tilde{<}$ with : $$( i', \zeta' , \epsilon' ,
\alpha' ) \quad \tilde{<} \quad ( i , \zeta , \epsilon , \alpha  )
$$ if and only if $$ i' =  i, \zeta' < \zeta , \epsilon' <
\epsilon , \alpha' < \alpha $$  This defines the neighborhoods
$\tilde{B}_{n}(g)$ as:
$$ \tilde{B}_{n} (g) = \{ \tilde{ g} | ( g , \tilde{g} ) \quad  \text{are}
\quad (  i', \zeta' , \epsilon' , \alpha' ) \quad \text{close
with} \quad (  i', \zeta' ,  \epsilon' , \alpha' ) \quad \tilde{<}
\quad ( n, \frac{1}{n} ,  \frac{1}{n} , \frac{1}{n}  ) \} $$ It is
easy to see that the topology defined by the sets
$\tilde{B}_{n}(g)$ is
the same as the one defined by the $B_{n}(g)$. \\
Now proposition $6$ again defines a pseudodistance on $ \mathbb{L}
( M ) $ as before, but this pseudodistance is only diagonally
invariant with respect to diffeomorphisms $\phi$ such that $
\phi_{ | W_{1}} \in Diff(W_{1})$.  So it seems we have to
symmetrize twice, thus we have to prove that $ f_{g, \tilde{g}} $
and $ \psi \rightarrow A(f_{ \psi^{*} g ,\tilde{g} }) $ are
continuous in the $ \mathcal{F} \mathcal{D}  $ topology. Choose $
\epsilon > 0$, then one can bound $$| A(f_{ \psi^{*}g , \tilde{g}}
) - A(f_{g , \tilde{g}}) |$$ by $$ {
\parallel f_{ \psi^{*}g , \tilde{g}} - f_{ g , \tilde{g}} \parallel}_{ \infty} $$ This can
again be bounded by $ d( g , \psi^{*} g ) $ because of the
triangle inequality. Now there exists an $n_{0}$ such that $
\psi^{*} g \in B_{ n_{0}} ( g ) $ implies that $ \psi^{*} g \in
B_{ \epsilon }^{d} ( g ) $. It is obvious now that because $
\bar{W}_{n_{0}}$ is compact, a similar result of proposition $8$
for $ W_{n_{0}}$ implies our result.  So we are only left with
proving that :
\begin{theo}
For any $i \in \mathbb{N}_{0}$ and $g , \tilde{g }$ globally
hyperbolic on $M$, we have that \begin{itemize} \item $ \phi
\rightarrow d^{i}_{\text{cau}}( g , \phi^{*} \tilde{g} )$
\item $ \phi
\rightarrow d^{i}_{\text{vol}}( g , \phi^{*} \tilde{g} )$
\item $ \phi
\rightarrow d^{i}_{\text{geo}}( g , \phi^{*} \tilde{g} )$
\end{itemize}
are continuous in the $ \mathcal{F} \mathcal{D} $ topology.
\\
Proof
\end{theo}
One has to keep in mind that $ \bar{W}_{i} $ is not necessarily a
compact submanifold of $M$.  So one cannot assume a differential
structure on $ \bar{W}_{i} $.  Let the $U_{i} $ be a countable
cover of charts of $M$ as in section $5.1$ and assume $U_{k}$ is a
finite subcover of $ \bar{W}_{i} $ then the conditions on the
determinant of $\phi$ obtained in proposition $11$ are meant with
respect to the finite subcover $U_{k}$.  This determines again
open neighborhoods of the identity in $ Diff(M)$ and we are done.
$\square$
\\
Now it is possible as in the compact case to introduce the
following pseudodistance:
$$ \tilde{d} : ( g , \tilde{g} ) \rightarrow \begin{cases}
A( \psi \rightarrow A( f_{\psi^{*}g ,\tilde{g}} + f_{
\psi^{*}\tilde{g} , g })) \quad \text{if and only if} \quad
\phi^{*} g \neq
\tilde{g} \quad \forall \phi \in Diff(M) \\
0 \quad \text{otherwise}
\end{cases} $$
The symmetrization would be unnecessary if an analogue to the
Fubini theorem would be valid for $A$. We don't know if this is
the case so we symmetrize anyway.  The mean $A$ in the above
construction is not necessarily an IM, it is easy to check that if
$A$ is a LIM then $\tilde{d}$ satisfies all properties of a
pseudometric. There is one disappointing feature to this
symmetrization, namely that we have been unable to prove that the
resulting topology does not depend on the choice of the
pseudometric which generates the uniformity.  We give an example
which illustrates this.
\begin{exie}
We consider $1+1$ dimensional space time $ \mathbb{R}^{2} $ and
the subspace $V \subset \mathbb{L} ( M ) $ defined as follows:
$$ V = \{ \begin{pmatrix} - \alpha & 0 \\ 0 &  \beta \end{pmatrix}
| \alpha , \beta > 0 \} $$ Define the action of $G =
\mathbb{R}^{+}_{0} \subset Diff( \mathbb{R}^{2}) $ on
$\mathbb{R}^{2}$ as follows:
$$ \breve{\lambda} ( x ,y ) = ( x , \sqrt{ \lambda } y ) \quad
\forall \lambda \in G $$  The $ \mathcal{F} \mathcal{D} $ topology
on $Diff(M)$ induces the discrete topology on $G$.  This is
obvious because if $ \breve{\lambda}, \breve{\gamma} $ are two
diffeomorphisms in $G$ which differ only on a set of compact
closure then they must be equal.  This topological group however
is not amenable.  We don't need such a strong topology to make the
functions $ \phi \rightarrow d^{i}_{\text{cau}}( g , \phi^{*}
\tilde{g} )$, $\phi \rightarrow d^{i}_{\text{vol}}( g , \phi^{*}
\tilde{g} )$ and $ \phi \rightarrow d^{i}_{\text{geo}}( g ,
\phi^{*} \tilde{g} )$ continuous ( the $W_{i}$ are circles of
radius $i$ in the standard Riemannian metric on $ \mathbb{R}^{2}
$).  They will also be continuous if we relax the discrete
topology to the standard euclidian one. The action of $G$ on $V$
is as follows:
$$ { \breve{ \lambda}}^{*} \begin{pmatrix} - \alpha & 0 \\ 0 &  \beta
\end{pmatrix} = \begin{pmatrix} - \alpha & 0 \\ 0 &  \lambda \beta
\end{pmatrix}$$
It is clear that the topology on $V \sim \mathbb{R}^{+}_{0} \times
\mathbb{R}^{+}_{0} $ defined by the uniform neighborhood system
$B_{  i , \zeta ,  \epsilon , \alpha } ( g ) $ is equivalent to
the standard euclidian topology on $\mathbb{R}^{+}_{0} \times
\mathbb{R}^{+}_{0} $.  Define on $ \mathbb{R}^{+}_{0} \times
\mathbb{R}^{+}_{0} $ the following two metrics ( with cut off
scale $ \eta $ ):
$$
d_{1} ( \bar{a} , \bar{b} )  =  \min \{ \parallel \bar{a} -
\bar{b}
\parallel , \eta \} $$
and $$ d_{2} ( \bar{a} , \bar{b} )  = \min \{ | a_{1} - b_{1} | +
| \frac{1}{a_{2}} - \frac{1}{b_{2}} | , \eta \}
$$
It is obvious that $d_{1}$ and $d_{2}$ generate the euclidian
topology, they are however not equivalent.  If we calculate :
$$\tilde{d}_{1}( \bar{a} , \bar{b} ) =  m( x \rightarrow m( y \rightarrow d_{1} ( (a_{1} , x a_{2} ), (
b_{1} , y b_{2} )))) $$ then one gets that $$
\tilde{d}_{1}(\bar{a} , \bar{b} ) = \begin{cases} \eta \quad
\text{ if } \quad \bar{a} \neq { \breve{ \lambda }}^{*}
\bar{b} \quad \forall \lambda \in G \\
0 \quad \text{otherwise} \end{cases} $$ The same calculation for
$d_{2} $ gives however a totally different result:  $$
\tilde{d}_{2}(\bar{a} , \bar{b} ) = \begin{cases} \min \{| a_{1} -
b_{1} | , \eta \} \quad \text{ if } \quad \bar{a} \neq { \breve{
\lambda }}^{*}
\bar{b} \quad \forall \lambda \in G \\
0 \quad \text{otherwise} \end{cases} $$ So in the first case the
quotient topology on $ V/G$ is the discrete topology, in the
second case it is the usual euclidian one.  This shows how
sensitive the quotient topology is to the choice of the generating
distance for the uniformity.
\end{exie}
This problem becomes even much more complicated when we consider
diagonally invariant (with respect to a group action $G$) metrics.
One can however make the following remark:
\begin{exie}
Consider the action of $ \mathbb{Q}^{+}_{0} $ by multiplication on
$ \mathbb{R}^{+}_{0} $ endowed with the metric $$ d(a,b) = \mid ln
\left( \frac{ a}{b} \right) \mid $$ It is clear that $ \sqrt{2} $
is not on the orbit of $ \sqrt{3} $ but this orbit is dense in $
\mathbb{R}^{+}_{0} $.
\end{exie}
This example is quite pathologic in the sense that the group $
\mathbb{Q}^{+}_{0}$ is totally disconnected.  This is however not
the case for $Diff(M)$.
\\

\section{Epilogue}
We will first compare our results with the results obtained in
~\cite{Bombelli,Bombelli2}.
\begin{itemize}
\item The topology on $ \mathbb{L}(M)/ Diff(M) $ constructed in ~\cite{Bombelli} is unique,
however there exist many pseudodistances which might generate this
uniform topology.  The topology on $ \mathbb{L} ( M ) / G $ in our
case is probably not uniquely determined and depends on the
generating pseudodistance of the uniform topology on $
\mathbb{L}(M)$.  However as mentioned in the introduction when $M$
is compact we can also take the quotient $ \mathbb{L}(M)/ Diff(M)
$.
\item Obviously  we don't know the "size" yet of the maximal amenable
subgroup $G$, this is something which has to be investigated.
\item Our topology has much better continuity properties with
respect to group actions.
\item One can also raise the question whether
the topology on $ \mathbb{L} (M) $ is locally arcwise connected.
$Diff(M)$ is locally arcwise connected in the $ \mathcal{F}
\mathcal{D} $ topology so we have that for $ \phi $ sufficiently
small and $ g \in \mathbb{L} ( M)  $ that $ \phi^{*} g $ is
arcwise connected to $g$ by a path in $ \mathbb{L} (M ) $ which
corresponds to a path in $ Diff( M) $ from the identity to $ \phi
$.
\end{itemize}
Let $d$ be the distance generating the uniform topology (on the
space of globally hyperbolic metrics $\mathbb{GLH}(M)$).  If we
would take the quotient $\mathbb{GLH}(M)/Diff(M)$, the Hausdorff
property would fail if and only if:
$$ \inf_{\psi \in Diff(M)} d(g, \psi^{*}\tilde{g}) = 0 $$ for $
\tilde{g} $ not diffeomorphism equivalent to $g$.  This implies
that there exists a sequence of diffeomophisms $\psi_{n}$ such
that $ \psi_{n}^{*} \tilde{g} \in B_{n} (g)$.  This means that for
$n$ big enough the conformal and volume structures of
$\psi_{n}^{*} \tilde{g}$ and $g$ are almost the same.  There is no
argument to exclude that this can happen.  By symmetrizing $d$
with an invariant mean we hope that this will happen less, however
there is no
way we can prove this. \\
We will show now that our work has hope to further generalization.
 As mentioned before, one is interested in measuring the distance
between isometry classes of Lorentzian structures $(M,g)$ and
$(M',g')$. $(M,g)$ and $(M',g')$ are isometrically equivalent if
and only if there exists a diffeomorphism $ \phi: M \rightarrow
M'$ such that $\phi_{*}g = g'$.  Suppose that all manifolds
considered are compact, and let $d_{M}$ be the (diagonally
invariant) distance defined in proposition $9$.  We define the
pseudodistance $\hat{d}$ between $(M,g)$ and $(M',g')$ as follows:
$$\hat{d}((M,g),(M',g')) = \begin{cases} \inf_{ \psi \in
Diff(M,M')} d_{M} (g, \psi^{*}g') \text{ if } Diff(M,M') \neq
\emptyset \\ 1 \text{ otherwise}
\end{cases} $$ The symmetry of $\hat{d}$ follows from $d_{M} (g,
\psi^{*}g') = d_{M'} ( \psi_{*} g, g' ) = d_{M'} (g' , \psi_{*} g
)$.  The triangle inequality follows from:
\begin{eqnarray*}
 d_{M} (g, \psi^{*}g'')  & \leq &
 d_{M}( g , \phi^{*} g' ) + d_{M} (\phi^{*} g' , \psi^{*} g'' )
  \\ & \leq & d_{M}( g , \phi^{*} g') + d_{M'}( g' ,
 ( \psi \phi^{-1} )^{*} g'' ) \\
 \end{eqnarray*}
$ \hat{d} $ is clearly a pseudodistance on isometry classes of
metrics.  Unfortunately $ \hat{d}$ makes a too rough distinction
between two inequivalent differential structures.  We propose now
a la Gromov a pseudodistance on the space of all compact, future
and past distinguishing structures $(M,g)$, which might improve
drastically this defect. We denote by $\mathcal{C}(M,M')$ the
space of all continuous functions from $M$ to $M'$.  We define two
structures $(M,g)$ and $(M',g')$ to be $(\zeta ,\epsilon , \mu )$
close if and only if there exist $f \in \mathcal{C}(M,M')$ and $ h
\in \mathcal{C}(M',M)$ such that:
\begin{itemize}
\item $ \sup_{ \mathcal{O} \in \mathcal{B}_{M}} \mid ln \left(
\frac{V_{g'} (\overline{f( \mathcal{O})})}{V_{g} ( \mathcal{O})}
\right) \mid < \epsilon $ and $ \sup_{ \mathcal{O} \in
\mathcal{B}_{M'}} \mid ln \left( \frac{V_{g} ( \overline{h(
\mathcal{O})})}{V_{g'} ( \mathcal{O})} \right) \mid < \epsilon $
where $\mathcal{B}_{M}$ and $\mathcal{B}_{M'}$ are the sets of all
Borel measurable subsets of nonzero measure of $M$ and $M'$
respectively.
\item $$ \sup_{p,q \in M: V_{g'}(\overline{f(A_{g}(p,q))} \cup A_{g'}(f(p),f(q))) \geq \zeta }
\frac{V_{g'}( \overline{f(A_{g}(p,q))} \bigtriangleup
A_{g'}(f(p),f(q))) }{ V_{g'}(\overline{f(A_{g}(p,q))} \cup
A_{g'}(f(p),f(q)))} \leq \epsilon $$ and a similar constraint with
$f$ replaced by $h$ and $(M,g)$ switched with $(M',g')$
\item We are also obliged to make a topological constraint in
order to construct a uniformity.  Intuitively this constraint
means that $h$ is approximately the inverse of $f$ and vice versa:
$$ \forall A \in \tau_{M} \text{ such that } V_{g} (A) \geq \mu : V_{g} ( A \bigtriangleup
\overline{h(f(A))})\leq \epsilon V_{g}(A)$$ and
$$ \forall A \in \tau_{M'} \text{ such that } V_{g'} (A) \geq \mu: V_{g'} ( A \bigtriangleup
\overline{f(h(A))})\leq \epsilon V_{g'}(A) $$ where $\tau_{M}$ (
$\tau_{M'}$) is the set of all closed subsets of $M$ ($M'$).
\end{itemize}
This notion of closeness defines again a uniformity, and hence a
pseudodistance $d$ on the space of all isometry classes of
compact, future and past distinguishing structures.  The reader
might ask why we go back to the supremum function while the
natural thing to do would be to introduce the following type of
constraint on the space of compact, class $ \mathcal{A}$
structures:
$$ \int_{M \times M} \frac{V_{g'}( \overline{f(A_{g}(p,q))}
\bigtriangleup A_{g'}(f(p),f(q))) }{
V_{g'}(\overline{f(A_{g}(p,q))} \cup A_{g'}(f(p),f(q)))}
dV_{g}(p)dV_{g}(q) < \epsilon$$ This seems also possible to us,
but one has to demand then that the functions $f$ and $h$ are
injective.

\section{Acknowledgement}
The author wishes to thank Prof. Norbert Van den Bergh for the
many discussions and his birds eye reading of the original
manuscript.  He also wants to express his gratitude to his
colleagues at the institute for the nice working atmosphere .

\newpage
\section{Appendix A}
In this appendix we introduce the class $\mathcal{A} $.  In
proposition $4$ it will be proven that $ \mathcal{A}$ contains all
globally hyperbolic metrics. In ~\cite{Szabados} one defined with
respect to a Lorentz metric $g$ a semiring $ \mathcal{H}_{g} $
with identity element as follows:
$$ \mathcal{H}_{g} = \{ F \cap P \mid F \quad \text{is a future set
and} \quad P \quad \text{is a past set} \} $$ It is also proven in
~\cite{Szabados} that all sets in $ \mathcal{H}_{g}  $ are
Lebesgue measurable.  Moreover we have that for any element $H \in
\mathcal{H}_{g} $ :
$$ \mu ( H ) = \mu ( \overline{H} ) = \mu ( \overset{ \circ}{H} )
$$
where $ \mu$ is the four-dimensional Lebesgue measure.
$\mathcal{H}_{g}$ is not closed with respect to the union and has
the property that the complement of an element  $ ( F \cap P )^{c}
= ( F^{c} \cup P^{c} )$ equals the union of two elements of $
\mathcal{H}_{g} $.  Denote by $ \mathcal{L} $ the $ \sigma $
algebra of all Lebesgue measurable sets, we define two topologies
on $ \mathcal{L} $ by giving a base. Let $ A $ be an open subset
and $ C $ a closed subset of $M$ satisfying the property that $
\overline{ \overset{ \circ }{C} } = C $ , the open set $B( A ; C)
$ is defined as follows:
$$ S \in B( A ; C ) \longleftrightarrow \overset{\circ}{C} \subset
\overset{\circ}{S} \subset A
$$ It is easy to check that these sets constitute a basis for the
topology $\tau_{1}$ on $ \mathcal{L} $.  The topology $ \tau_{2} $
is defined in a slightly different way: the open sets $B(A,C) $
constitute of the following elements: $$ S \in B(A,C )
\longleftrightarrow C \subset \overset{ \circ}{S} \subset
\overline{S} \subset O $$ This leads us to the following
definition:
\begin{deffie}
A Lorentz metric $g$ is of class $ \mathcal{A}$  if and only if
the map:
$$ A: M \times M \rightarrow \mathcal{H}_{g} : (p,q) \rightarrow
A(p,q) $$ is measurable with respect to one of the Borel $\sigma$
- algebras defined by the topological spaces $ ( \mathcal{H}_{g} ,
\tau_{1} )$, $ ( \mathcal{H}_{g} , \tau_{2} )$ .
\end{deffie}
In a next paper we will examine class $ \mathcal{A} $ space times
more thoroughly.  From now on it is always assumed that the space
time metrics are of class $\mathcal{A} $.  It is not difficult to
prove that the map $ \alpha_{g \tilde{g} } $ is measurable with
respect to class $ \mathcal{A} $ space times.  One can also show
that on a compact space time a causally continuous metric is of
class $ \mathcal{A} $, one even has the stronger result that the
corresponding map $A$ is continuous with respect to the topology $
\tau_{2} $.

\section{Appendix B}
We have to prove that for any $ \epsilon > 0 , \quad \exists
\delta_{1} > 0 $ such that
$$ V^{2}( \bigcup_{p \in M} S^{ \delta_{1} }_{p} ) \leq \frac{\epsilon}{2}
$$
\\
Choose $ \epsilon > 0 $, because $ \delta \rightarrow V( \{ p |
V(\tilde{J}^{+} ( p )) \leq \delta \} ) $ is continuous,
monotonously increasing and $ V(\{ p | V(\tilde{J}^{+} ( p )) =0
\} ) = 0 $, we can find a $ \delta_{2} < 1 $ such that $ V( \{ p |
V(\tilde{J}^{+} ( p )) \leq \delta_{2} \} )  < \frac{ \epsilon
}{4} $. We denote in the sequel $\gamma = \frac{\epsilon}{ 4
\tilde{M} }, \quad V(M) = \tilde{M}, \quad M^{ \delta_{2}} = \{ p
| V(\tilde{J}^{+} ( p )) \geq \delta_{2} \} $ and define $ \forall
p \in M^{ \delta_{2}} $ the mapping $$ \eta_{p} : [ 0 , \tilde{M}
] \rightarrow \mathbb{R} : x \rightarrow V( S^{x}_{p})
$$
It is not difficult to see that the mappings $ \beta_{p}, \eta_{p}
$ are continuous on $M$ and nonzero on $ M^{ \delta_{2}}$,
moreover $ \forall p \in M^{ \delta_{2}} $ there exists a unique
$\delta_{p} > 0 $ such that the mapping $ \eta_{p} $ is strictly
increasing for $ x < \delta_{p} $ and constant for $ x \geq
\delta_{p} $.  It is obvious that $ \forall p \in M^{ \delta_{2}}
$ there exists a unique $ \tilde{\delta}^{ \gamma }_{p}
> 0 $ such that that $ x
> \tilde{\delta}^{ \gamma }_{p} $ implies that $ \eta_{p}( x ) >
\gamma $ or $ x > \delta_{p} $. If we prove now that the mapping
$p \rightarrow \tilde{\delta}^{ \gamma }_{p}$ is continuous then
we are done because $ \forall \quad  0 < \nu \leq \min_{p \in M^{
\delta_{2}}} \tilde{\delta}^{ \gamma }_{p} = \delta_{1} $ we have
that $ \eta_{p} ( \nu ) \leq \gamma \quad \forall p \in M^{
\delta_{2}}$ and so $$ V^{2}( \bigcup_{p \in M} S^{ \nu }_{p} ) <
\gamma \tilde{M} + \frac{ \epsilon}{ 4 } < \frac{ \epsilon }{2} $$
\\
To prove the continuity of $p \rightarrow \tilde{\delta}^{ \gamma
}_{p}$, we first show that the mapping $$ M \times [ 0 , \tilde{M}
] \rightarrow \mathbb{R} : ( p , \nu ) \rightarrow \eta_{p}( \nu )
$$ is continuous.  Choose $ \tilde{ \epsilon } > 0 $ and $
( p , \nu ) \in M \times [ 0 , \tilde{M} ] $.  Because $ \eta_{p}
$ is uniformly continuous there exists a $ \tilde{\delta}_{1} $
such that $ | x - y | < \tilde{\delta}_{1} $ implies that $ |
\eta_{p} ( x ) - \eta_{p} ( y ) | < \frac{ \tilde{ \epsilon }}{ 4}
$. Then there exists a $ \tilde{ \delta }_{2} > 0 $ such that $
\forall q \in B( p , \tilde{ \delta }_{2} ) : V( \tilde{J}^{+}( p)
\bigtriangleup \tilde{J}^{+} (q)) < \frac{ \tilde{ \epsilon }}{2}
$ and $ | \beta_{p}(r) - \beta_{q} (r) | <
\frac{\tilde{\delta}_{1}}{2} \quad \forall r \in \tilde{J}^{+} (
p)$ because $\tilde{J}^{+} ( p)$ is compact.  This implies that
$$ | \eta_{p} ( \nu ) - \eta_{q} ( \chi ) | < V( \{ r \in
\tilde{J}^{+}(p) \cap \tilde{J}^{+} ( q ) | \beta_{p} ( r) > \nu
\quad \text{or} \quad \beta_{q} ( r) > \chi \quad \text{ but not
both } \} ) + \frac{ \tilde{ \epsilon }}{2} $$ This is smaller
than $$ V( \{ r \in \tilde{J}^{+}(p) \cap \tilde{J}^{+} ( q ) |
\nu < \beta_{p} ( r) \leq \chi + \frac{\tilde{ \delta}_{1}}{2}
\quad \text{and} \quad \beta_{q} ( r) \leq \chi \} )+ $$ $$V( \{ r
\in \tilde{J}^{+}(p) \cap \tilde{J}^{+} ( q ) | \chi -
\frac{\tilde{ \delta}_{1}}{2} < \beta_{p} ( r) \leq \nu \quad
\text{and} \quad \beta_{q} ( r) > \chi \} )+\frac{ \tilde{
\epsilon }}{2} $$ and this can be further bounded by $$ | \eta_{p}
( \chi + \frac{\tilde{ \delta}_{1}}{2} ) - \eta_{p} ( \nu ) | + |
\eta_{p} ( \nu ) - \eta_{p} ( \chi - \frac{\tilde{ \delta}_{1}}{2}
) | + \frac{ \tilde{ \epsilon }}{2} $$ which is smaller than $
\tilde{ \epsilon } $ for all $ \chi , \nu $ such that $ | \chi -
\nu | < \frac{\tilde{ \delta}_{1}}{2} $. \\ Now to prove the
continuity of $p \rightarrow \tilde{\delta}^{ \gamma }_{p}$ we
have to consider two cases:
\begin{itemize}
\item $ \tilde{ \delta }^{ \gamma }_{p} < \tilde{ \delta }_{p} $.
Because $ \eta_{p} $ is strictly increasing in $ \tilde{ \delta
}^{ \gamma }_{p} $ there exists for $ \kappa > 0 $ small enough a
unique $ \zeta > 0 $ such that $ | \eta_{p} ( \nu ) - \eta_{p} (
\tilde{ \delta }^{ \gamma }_{p} ) | < \zeta $ implies that $ | \nu
- \tilde{ \delta }^{ \gamma }_{p} | < \kappa $. Choose now $
\delta > 0 $ small enough such that $ q \in B(p , \delta )$
implies that $$| \eta_{p} ( \nu ) - \eta_{q} ( \nu )| < \zeta
\quad \forall \nu \in [ 0 , \tilde{M} ] $$ ( this is possible
because of the continuity of $ ( p , \nu ) \rightarrow \eta_{p}(
\nu ) $).  We obviously have then that $  | \tilde{ \delta }^{
\gamma }_{p} - \tilde{ \delta }^{ \gamma }_{q} | < \kappa \quad
\forall q \in B(p , \delta )$.
\item We are left with the case $ \tilde{ \delta }^{ \gamma }_{p}  = \tilde{ \delta }_{p} $.
Here the result will immediately follow from the previous result
and the fact that $ \forall \kappa > 0 \quad \exists \delta
> 0$ such that $ \forall q \in B(p, \delta ) :  \beta_{q} (r)  < \tilde{ \delta }_{p}
+ \kappa \quad \forall r \in \tilde{J}^{+} (q) $.  Suppose this is
false then $ \exists q_{n} \prec q_{m} \prec p \quad \forall n < m
$ and a sequence $ r_{n} \in \tilde{J}^{+} (q_{n}) : \beta_{q_{n}
} (r_{n} ) \geq \tilde{ \delta }_{p} + \kappa $. Because $M$ is
compact the sequence $ r_{n} $ has an accumulation point $r$.
Clearly $ r \in \tilde{J}^{+} (p)$ and $ \beta_{p } (r ) \geq
\tilde{ \delta }_{p} + \kappa $ which is impossible.
\end{itemize}
$\square$
\section{Appendix C}
Here we prove lemma 1.  One has to show that the function $$ F :
M^{2}_{ \delta_{1}} \rightarrow \mathbb{R} : ((p,q),(r,s))
\rightarrow \frac{ V( \tilde{A}( p,q) \bigtriangleup \tilde{A}( r
,s ))}{ V( \tilde{A} (p,q)) } $$ $F$ is continuous.  \\
Proof \\ Choose $ 1 > \kappa > 0 $, then one easily sees that $ V(
\tilde{A}( p',q') \bigtriangleup \tilde{A}( r' ,s' ))$ can be
bounded by:
$$ V( \tilde{A}( p,q)
\bigtriangleup \tilde{A}( p' ,q' ))+V( \tilde{A}( p,q)
\bigtriangleup \tilde{A}( r ,s ))+V( \tilde{A}( r',s')
\bigtriangleup \tilde{A}( r ,s ))$$ Now obviously there exists a $
\delta > 0$ such that $ p' \in B(p, \delta ),  q' \in B(q, \delta
),  r' \in B(r, \delta ),  s' \in B(s, \delta )$ implies that with
$ \nu = \frac{ \kappa \delta_{1} }{ 3 \delta_{1} + \tilde{M}} $ we
have that
$$ V( \tilde{A}( p,q) \bigtriangleup \tilde{A}( p' ,q' )) <  \nu V( \tilde{A}( p,q))$$ and
$$ V( \tilde{A}( r,s) \bigtriangleup \tilde{A}( r' ,s' )) <  \nu V( \tilde{A}( p,q)) $$ This implies
that $$ F(p',q',r',s') < (\frac{1}{ 1 - \nu} )(  F(p,q,r,s) + \nu
\frac{ \delta_{1} + \tilde{M} }{ \delta_{1}} )< F(p,q,r,s) +
\kappa
$$
The lower continuity is proven in a similar way. \\ $\square $

\section{Appendix D}
We prove the following result:
\\
let $ \tilde{ \epsilon } > 0 $, then there exists a neighborhood
$U_{ \tilde{ \epsilon }}$ of the zero section such that $ \forall
(p,q) \in M_{ \delta_{1}} $ we have that  :
$$ V(\tilde{A} (\phi(p),\phi(q)) \bigtriangleup \phi^{-1}
\tilde{A} ( \phi (p) , \phi (q) ) ) < \tilde{ \epsilon }V(
\tilde{A} ( p ,q ))
$$ $ \phi \in U_{ \tilde{ \epsilon }}$ implies that $ d_{vol} ( \phi_{*} g , g ) < ln(3) $.
\\
Proof
\\
From the previous lemma we can assume that for every $ \phi \in
U_{ \tilde{ \epsilon }}$ we have that $ V(\tilde{A}
(\phi(p),\phi(q)) \bigtriangleup \tilde{A} ( p , q) ) <  \frac{
\tilde{\epsilon} }{ 8}  V( \tilde{A}( p,q))$. It is easy to see
that $ V(\tilde{A} (\phi(p),\phi(q)) \bigtriangleup \phi^{-1}
\tilde{A} ( \phi (p) , \phi (q) ) ) $ can be bounded by
$$ V(\tilde{A} (\phi(p),\phi(q)) \bigtriangleup
\tilde{A} ( p , q) )  +  V(\tilde{A} (p,q) \bigtriangleup
\phi^{-1} \tilde{A} ( p , q )) + V(\phi^{-1}\tilde{A}
(\phi(p),\phi(q)) \bigtriangleup \phi^{-1} \tilde{A} ( p , q) ) $$
The last term is equal to $$ \phi_{*} V( \tilde{A}
(\phi(p),\phi(q)) \bigtriangleup \tilde{A} ( p , q) ) < \frac{ 3
\tilde{ \epsilon }}{8} V( \tilde{A}( p,q))
$$ Thus everything can be bounded by:
$$ \frac{
\tilde{ \epsilon }}{2 } V(\tilde{A} ( p , q) ) + V(\tilde{A} (p,q)
\bigtriangleup \phi^{-1} \tilde{A} ( p , q ))  $$ and we are left
to prove that there exists an open neighborhood of the identity
such that: $$ \phi_{*} V ( \tilde{A} (p,q) \bigtriangleup \phi
\tilde{A} ( p , q )) = V(\tilde{A} (p,q) \bigtriangleup \phi^{-1}
\tilde{A} ( p , q )) < \frac{\tilde{ \epsilon }}{2 } V(\tilde{A} (
p , q) )$$ $ \forall (p,q) \in M_{ \delta_{1}} $. It is clearly
possible to find an open neighborhood $V$ of the identity such
that in one point $(p,q)$ one has:
$$ V(\tilde{A} (p,q) \bigtriangleup \phi \tilde{A} ( p , q ))
< \frac{\tilde{ \epsilon }}{12 } V(\tilde{A} ( p , q) )$$  Now it
is also obvious that the mapping $$ (r,s) \rightarrow
\frac{V(\tilde{A} (r,s) \bigtriangleup \phi \tilde{A} ( r , s ))}{
V(\tilde{A} ( r , s) )} $$ is continuous in $(p,q)$ using similar
techniques as above.  So we can find a $ \delta > 0 $ such that $
\forall r \in B(p , \delta ), s \in B( q , \delta ) $ we have that
$$ V(\tilde{A} (r,s) \bigtriangleup \phi \tilde{A} ( r , s ))
< \frac{\tilde{ \epsilon }}{6 } V(\tilde{A} ( r , s) )$$   The
compactness of $ M_{ \delta_{1}} $ together with $ \phi_{*} V \leq
3 V $ yield the result. $ \square $
\section{Appendix E}
Let $G$ be a locally compact group, $ \lambda $ the right
invariant Haar measure and $ \Delta $ the (right) modular function
ie. $ \lambda ( f^{g} ) = \Delta(g) \lambda ( f )$ where $
f^{g}(x) = f(g^{-1} x ) $.  A l.c. group is unimodular if and only
if the modular function equals $1$.  We denote with $ K(G) $ the
functions of compact support, $P^{p}(G) $ are the positive
elements of $ L^{p} (G) $, $L^{ \infty }(G)$ has here the meaning
of the essentially bounded measurable functions ( the locally null
sets are defined with respect to $ \lambda $ ) and $M^{1} ( G ) $
denotes the convex set of the probability measures on $G$.  We
denote with $f*g$ the usual convolution product.  The following
few properties characterize amenability:
\begin{itemize}

\item There exists a net $ ( \phi_{i} )$ in $P^{p} ( G ) \cap K(G) $
such that $$ \lim_{i} {\parallel \phi^{a}_{i} - \phi_{i}
\parallel}_{p} = 0 $$ whenever $ a \in G$

\item There exists a net $ ( \phi_{i} )$ in $P^{1} ( G) $ such that
$$ \lim_{ i } { \parallel \chi * \phi_{i} - \phi_{i} \parallel
}_{1} =0 $$ whenever $ \chi \in M^{1} ( G) $.

\item If $K$ is a compact subset (resp. $F$ is a finite subset)
and $ \epsilon > 0 $ then there exists a measurable set $U$ in $G$
of positive finite measure such that $$ \frac{\lambda ( a U
\bigtriangleup U )}{ \lambda ( U ) } \leq \epsilon \quad \forall a
\in K \quad ( a \in F ) $$

\item  If $K$ is a compact subset (resp. $F$ is a finite subset)
and $ \epsilon > 0$ then there exists a measurable set $U$ of
positive finite measure such that :
$$ \frac{ \lambda ( KU ) }{ \lambda ( U ) } < 1 + \epsilon $$ and
the same for the finite set.

\end{itemize}
One has of course much more characterizations of amenability; one
of them links amenability to properties of the continuous unitary
representations of $G$  ~\cite{Pier}.  There are special classes
of locally compact groups which are known to be (or not to be)
amenable. We give a grasp out of a wealth on results :
\\

\begin{itemize}

\item Every abelian l.c. group is amenable

\item If for every compact neighborhood of the identity in $G$ :
$$ \liminf_{ n \rightarrow \infty } \frac{\lambda ( V^{n+1}) }{ \lambda ( V^{n} )} = 1$$
then $G$ is amenable.  Notice also that the above condition
implies that $G$ is unimodular.  This result implies also that
every compact group is amenable.  Another consequence is that if
the conjugacy classes are relatively compact and $G$ is totally
disconnected then $G$ is amenable.

\item If $G$ is almost connected and semisimple, then $G$ is
amenable if and only if $G$ is compact

\item if $G$ is almost solvable then it is amenable.

\item If $G$ is a discrete free group that is freely generated by
elements $ a_{1} \ldots a_{n} $ of orders $ p_{i } $ then $G$ is
nonamenable unless $n=2$ and $ p_{1} = p_{2} = 2$.

\end{itemize}
One proves that any closed subgroup of an amenable l.c. group is
amenable, that the finite cartesian product of l.c amenable groups
is amenable.  So one easily concludes for example that a l.c.
group is not amenable if there exists a closed free subgroup of
$n$ generators where one generator has order different form $2$.

\end{document}